\documentclass[11pt]{article}
\usepackage{amsmath,amsthm,amssymb,amsfonts}
\usepackage{latexsym}
\usepackage{graphicx,psfrag,import}
\usepackage{fullpage}
\usepackage{framed}
\usepackage{verbatim}
\usepackage{color}
\usepackage{epsfig}
\usepackage{hyperref}
\usepackage{geometry}
\usepackage{mathtools}
\usepackage{float}
\usepackage{enumerate}
\usepackage{multicol}
\usepackage{a4wide}
\usepackage{booktabs}
\usepackage{enumitem}
\usepackage{lineno}
\usepackage{parcolumns}
\usepackage{thmtools}
\usepackage{authblk}
\usepackage{xr}
\usepackage{epstopdf}
\usepackage{mathrsfs}
\usepackage{subfigure}
\usepackage{caption}

\newsavebox{\measurebox}

\theoremstyle{plain}

\theoremstyle{definition}

\newcommand{\Int}{\int\limits}

\newcommand{\minus}{\scalebox{0.5}{$-$}}

\date{}

\renewcommand{\figurename}{FIGURE}

\begin{document}

\title{\textbf{High rates of  fuel consumption are not required by insulating motifs to suppress retroactivity in biochemical circuits}}
\author[1,2]{Abhishek Deshpande}
\author[3]{Thomas E. Ouldridge\thanks{Corresponding author: t.ouldridge@imperial.ac.uk}}
\affil[1]{Department of Mathematics, Imperial College London, London SW7 2AZ, United Kingdom}
\affil[2]{School of Technology and Computer Science, Tata Institute of Fundamental Research, Mumbai 400005, India}
\affil[3]{Department of Bioengineering, Imperial College London, London SW7 2AZ, United Kingdom}
\maketitle

\begin{abstract}
Retroactivity arises when the coupling of a molecular network $\mathcal{U}$ to a downstream network $\mathcal{D}$ results in signal propagation back from $\mathcal{D}$ to $\mathcal{U}$. The phenomenon represents a breakdown in modularity of biochemical circuits and hampers the rational design of complex functional networks. Considering simple models of signal-transduction architectures, we demonstrate the strong dependence of retroactivity on the properties of the upstream system, and explore the cost and efficacy of fuel-consuming insulating motifs that can mitigate retroactive effects. We find that simple insulating motifs can suppress retroactivity at a low fuel cost by coupling only weakly to the upstream system $\mathcal{U}$. However, this design approach reduces the signalling network's robustness to perturbations from leak reactions, and potentially compromises its ability to respond to rapidly-varying signals.
\end{abstract}

\maketitle

\begin{multicols}{2}

\section{Introduction}

The possibility of designing electrical circuits to function as well-defined modules is crucial to the engineering of complex circuitry with many interconnected components. Ideal modules have clearly-defined inputs, and produce outputs specified by those inputs and the internal structure of the module, independent of the broader context within which they are embedded~\cite{alon2007network}. This modularity property is also at the heart of computing, where it permits an extendable programming framework.

Biological systems appear to exhibit modularity in some contexts, and it has been suggested that this modularity is evolutionarily advantageous~\cite{Kashtan27092005,hintze2008evolution,Clune20122863,raff2000dissociability}, or contributes to system robustness~\cite{Tran20130771,rorick2011protein,kim2016robustness}. For humans seeking to engineer biological systems without the benefit of billions of years of highly parallel evolution, modularity can be hugely advantageous.
%  Indeed, synthetic biology is based on the principle that genes and even whole gene networks can be transplanted between cell lines and remain functional~\cite{purnick2009second}. 

\begin{figure*}[ht!]
\centering
\includegraphics[scale=0.16]{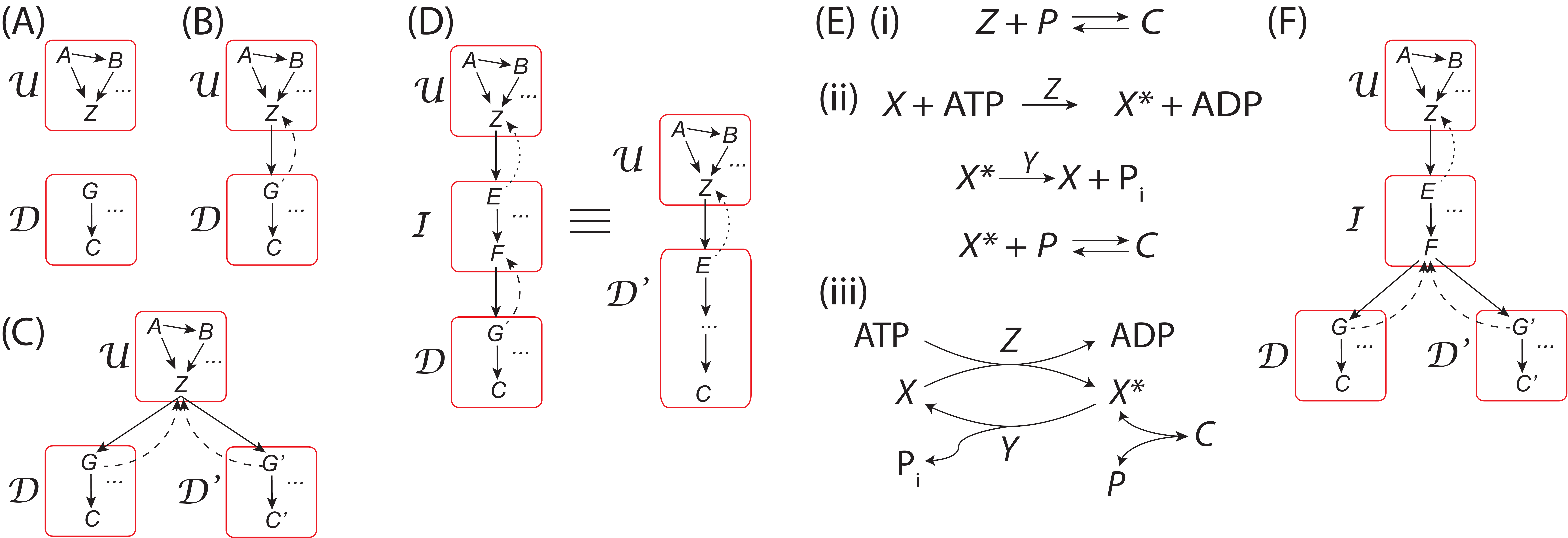}
\caption{A schematic representation of the concept of retroactivity. (A) Subsystems $\mathcal{U}$ and $\mathcal{D}$ evolve separately according to  internal dynamics  (the arrows in these diagrams indicate arbitrary interactions). (B) By coupling $\mathcal{U}$ and $\mathcal{D}$ via a molecular interaction between $Z$ (an output of $\mathcal{U}$) and ${G}$ (an input of $\mathcal{D}$), a signal is propagated. However, in general the coupling also induces changes in $Z$ due to retroactivity (dashed line). (C) This retroactivity is particularly problematic when adding an additional downstream system $\mathcal{D}^\prime$ that also couples to  $\mathcal{U}$. (D) Retroactivity can potentially be reduced using an insulator $\mathcal{I}$ between $\mathcal{U}$ and $\mathcal{D}$. In effect, $\mathcal{U}$ couples to a compound downstream motif $\mathcal{D}^\prime$. (E) A specific example of an insulating circuit. Retroactivty is very high for propagation of a signal by direct binding, as in (E.i). Alternatively, $Z$ can act as a catalyst for the phosphorylation of an intermediate species, $X\rightarrow X^*$, that when phosphorylated binds to $P$, thereby reducing the retroactivity experienced by $Z$, as in (E.ii). Note that this insulating ``push-pull" motif requires turnover of ATP to function, and involves an antagonistic phosphatase $Y$. (E.iii) A graphical representation of the reactions in (ii). (F) If the insulating circuit $\mathcal{I}$  itself couples to multiple downstream subsystems, retroactivity experienced by $\mathcal{I}$ may be relevant.}\label{fig:schematic}
\end{figure*} 

In the context of biochemical networks, the concept of modularity was first put on a solid footing by Hartwell~\cite{hartwell:molecular}, Lauffenburger~\cite{Lauffenburger09052000} and Weiss et.al.~\cite{purnick2009second,andrianantoandro2006synthetic}. The phenomenon of ``retroactivity" \cite{SaezRodriguez2005619,ventura2008hidden,ninfa,del2015biomolecular,sontag_barton}, illustrated formally in Fig.~\ref{fig:schematic}\,A,B has been shown to cause  a breakdown in modularity. Here, an upstream system $\mathcal U$ consisting of a set of molecular species and reactions is coupled to a downstream system $\mathcal{D}$ via a species $Z$, which is part of $\mathcal{U}$. The coupling causes a change in the output of $\mathcal{D}$ (here represented by a species $C$), passing on a signal from $\mathcal{U}$ to $\mathcal{D}$. However, in general $\mathcal{U}$  and in particular $Z$ is also affected by the coupling, implying the propagation of an unintended signal back from $\mathcal{D}$ to $\mathcal{U}$. Thus the meaning of the basic concepts of ``upstream" and ``downstream" is corrupted, and the ability to logically design circuits with well-defined inputs and outputs is compromised.  

The presence of retroactivity is particularly problematic when coupling to a subsystem $\mathcal{U}$ with pre-existing connections (Fig.~\ref{fig:schematic}.C) due to the fan-out effect~\cite{kim2010}. Such a situation could arise from a change in network topology from human intervention or natural evolution, or due to dynamic changes in molecular abundance within a cell. In these circumstances, strong retroactivity would lead to the new ``downstream" subsystem having an undesirable influence on the other ``downstream" subsystems, and vice-versa. 

Having identified the possibility of undesired retroactive interactions, several questions present themselves. Most immediately, how should retroactivity be quantified? Given a suitable metric, are certain designs of $\mathcal{U}$ and $\mathcal{D}$ more prone to retroactive effects? Is it possible to design insulating motifs $\mathcal{I}$, as illustrated in Fig.~\ref{fig:schematic}.D, that suppress retroactivity between $\mathcal{U}$ and $\mathcal{D}$? Does suppression of retroactivity necessarily imply an increased fuel consumption, and are there trade-offs associated with, for example, the accuracy of signal propagation?

In the last decade, several groups have considered these questions. In particular, Del Vecchio et.al.~\cite{ninfa,del2015biomolecular,shah2017signaling} have proposed the relative change in $Z$ due to the introduction of $\mathcal{D}$ as a potential metric for retroactivity. Later, Barton and Sontag~\cite{sontag_barton,barton_remark} proposed two alternative metrics to quantify retroactivity, namely the distortion and competition effect. The distortion captures the change in $C$ relative to an idealised system with no retroactive effect, while the competition effect quantifies the consequence for an existing downstream subsystem when a new one is attached to $U$. Particular attention has been paid to retroactivity in the context of the binding of transcription factors to DNA~\cite{ninfa,del2015biomolecular,ghaemi,sontag_barton,shah2017signaling,alon2006introduction}. Certainly, the passing of signals via binding is naturally retroactive, since it intrinsically requires sequestration of the upstream molecule. Moreover, signal propagation by direct binding occurs in contexts other than  transcription factor binding \cite{Lemmon1996}, and is widely used as a way to transmit signals in engineered  nucleic acid systems both {\it in vitro} and {\it in vivo} \cite{Qian2011,Green2014}. A recent review article by Hernandez and Garcia~\cite{pantoja2015retroactivity} gives a lucid account of the history of retroactivity and other works of a similar vein.

Sub-networks described as ``insulators" ($\mathcal{I}$) have been proposed to mitigate retroactivity by connecting $\mathcal{D}$ to $\mathcal{U}$ indirectly (see Fig.~\ref{fig:schematic}.D)~\cite{ninfa,sontag_barton,shah2017signaling,barton_remark}. The key component underlying these insulators is catalysis. Put simply, by acting as a catalyst molecule, the $Z$ is able to influence the downstream reactions without being sequestered indefinitely~\cite{Mehta2016,Ouldridge_PRX_2017}. 

The push-pull motif (or futile cycle) illustrated in Fig.~\ref{fig:schematic}~.E.ii is a common catalytic motif in natural signalling systems~\cite{robinson1997mitogen} and is also known by the name of ``futile'' cycle in literature~\cite{Samoilov15022005}. The molecule $X$ is catalytically switched between its two states $X$ and $X^*$ by the antagonistic enzymes $Z$ and $Y$ (respectively, a kinase and a phosphatase if the modification of $X$ is phosphorylation, as in Fig.~\ref{fig:schematic}.E.ii). The output of $X^*$ is then sensitive to the relative concentrations of $Z$ and $Y$. By this mechanism, a signal encoded in the concentration of $Z$ can be propagated without permanent binding of $Z$ to a downstream substrate. Previous work has shown that such a push-pull motif can function as an effective insulator between $\mathcal{U}$ and $\mathcal{D}$~\cite{ninfa,sontag_barton}, allowing information in the concentration of $Z$ to be propagated via $X^*$ to a downstream system with only limited sequestration of $Z$. 
Cascades of push-pull networks have also been considered~\cite{shah2017signaling,barton_remark,ossareh2011long,mishra2014load,sepulchre2012retroactive,catozzi2016signaling,wynn2011kinase,feliu2012algebraic}.

Catalysts cannot alter the equilibrium point of a reaction. Thus if $[X^*]$ is to be sensitive to $[Z]$, the system must be driven out of equilibrium by the turnover of biochemical fuel molecules~\cite{Mehta2016}. In the case of phosphorylation-based signalling, the system is driven out of equilibrium by the coupling of phosphorylation/dephosphorylation cycles to the breakdown of ATP into ADP and inorganic phosphate P$_{\rm i}$, as shown in Fig.~\ref{fig:schematic}.E.ii. The ATP molecules, which have a high free energy, are the chemical fuel. On a fundamental level, this fuel consumption (breakdown of ATP) allows the $X$ molecules to ``remember" the fact that they interacted with either $Y$ or $Z$ most recently, even though the interaction has ended~\cite{Govern09122014,Ouldridge_PRX_2017}.

For the catalytic reactions in Fig.~\ref{fig:schematic}.E.ii to proceed, complexes between $Z$ and $X$ must exist for a finite time~\cite{huang1996ultrasensitivity}. Therefore, although the push-pull insulator can reduce retroactivity, some of the $Z$ molecules are still sequestered at any given point and so some retroactivity remains. Barton and Sontag~\cite{sontag_barton} explored the question of whether this residual retroactivity could be suppressed, concluding that substantial energy consumption in the form of a high turnover of chemical fuel molecules was required. In a sequel~\cite{barton_remark}, they considered a slightly modified direct binding system in which the insulator itself acts catalytically on the downstream system $\mathcal{D}$, reaching the same conclusion. 

In this paper we revisit the resource costs of retroactivity suppression, as first highlighted in Ref~\cite{sontag_barton}, by considering the simplest steady-state setting. In Section~\ref{sec:direct_binding}, we consider whether the design of the upstream system $\mathcal{U}$ can mitigate retroactive effects when long-lived binding of $Z$ is  necessary for signal propagation. We observe that a constant turnover of $Z$ due to continuous production and decay can itself mitigate retroactivity in certain circumstances. However, this turnover is associated with a large resource cost if it is to be more rapid than the time-scale of signal variation. 

As a result, we turn our attention in Section~\ref{subsec:push_pull} to the analysis of insulating push-pull motifs that can potentially reduce retroactivity with limited protein production costs. Our main finding is that a higher rate of fuel consumption  is not required to produce better insulators. In general, both fuel consumption and retroactivity can be reduced simply by decreasing the coupling strength of both the upstream molecule $Z$ and the phosphatase $Y$ shown in Fig.~\ref{fig:schematic}E.ii to the $X/X^*$ molecule, whilst maintaining the steady-state output of the system. In Section~\ref{subsec:finite_free_energy}, we generalise this result to account for microscopic reversibility in the catalytic reactions that was neglected in Ref.~\cite{sontag_barton}. We find that a large chemical driving force (a large free energy stored in each ATP molecule) is necessary to propagate strong signals, but not to suppress retroactivity, and that it is still possible to reduce both retroactivity and energy consumption to low levels by reducing the coupling of $Z$ and $Y$ to the push-pull network.

Although high free-energy consumption is not necessary to suppress retroactivity, we postulate that a certain amount is indeed important for faithful signal transduction. In particular, in Section~\ref{sec:leak_reactions} we show that weak coupling of $Z$ and $Y$ to the push-pull network renders the system as a whole vulnerable to unintended leak reactions. We also hypothesize that high turnover of fuel molecules is necessary to accurately track time-dependent inputs to $\mathcal{U}$. 

\section{Methods}

We will work in the limit of a large copy number of molecules, in which case the reaction networks can be modelled deterministically by  mass-action ordinary differential equations. All our calculations related to retroactivity and energy consumption will be performed after the decay of initial transients, as in previous works~\cite{ninfa,sontag_barton,Sontag2011,pantoja2015retroactivity,jayanthi2013retroactivity}. Previously, some authors have focussed on systems driven by time-dependent variation of parameters within $\mathcal{U}$, such as birth and death rates of $Z$~\cite{ninfa,sontag_barton}. In our case, we will assume slow signal variation that implies that our systems reach steady states -- as in Ref.~\cite{ghaemi}. We make this simplifying assumption to deconvolve the distinct problems of tracking a time-dependent signal and suppressing retroactivity.

Taking our cue from Refs.~\cite{ninfa,del2015biomolecular,shah2017signaling} we use the retroactivity metric
\begin{eqnarray}\label{eq:retroactivity_metric}
\mathcal{R}= \left| 1 - \frac{[Z^{\rm ss}]}{[Z_{\mathcal{D,I} \rightarrow \emptyset}^{\rm ss}]}\right|,
\end{eqnarray}
where $[Z^{\rm ss}]$ denotes the concentration of free $Z$ in steady state in the actual system, $[Z_{\mathcal{D,I}  \rightarrow \emptyset}^{\rm ss}]$ is the steady state concentration of free $Z$ when it is not connected to the downstream system ($\mathcal{D}$ and $\mathcal{I}$ are absent, $\mathcal{D,I} \rightarrow \emptyset$). We find this metric to be more natural than others proposed~\cite{sontag_barton}, since it directly quantifies the back action on the upstream system. A more detailed analysis of the metrics used by other authors will form part of Section~\ref{sec:conclusion}. Throughout the manuscript, we will use the following reduced units: concentrations will be measured relative to $[C_0]= 10^{\minus 6}\rm M$. Uni-molecular rate constants will be measured relative to $k_0 = 1s^{-1}$ units and bimolecular rate constants relative to $\frac{k_0}{[C_0]}= 10^6\rm M^{-1}s^{-1}$. 

In addition to the retroactivity on $\mathcal{U}$, one could also consider the  retroactivity experienced by $\mathcal{I}$. This effect was termed ``retroactivity to the output" by Del Vecchio {\it et al.}~\cite{ninfa}, who considered the design of compound motifs for minimising both input and output retroactivity simultaneously \cite{ossareh2011long,shah2017signaling}. However, for a simple insulator  intended to prevent retroactivity on $\mathcal{U}$, any retroactivity experienced by $\mathcal{I}$ is irrelevant -- in the same way that the retroactivity internal to the compound motifs of 
Del Vecchio {\it et al.} is ignored \cite{shah2017signaling}. We therefore restrict our analysis to the retroactivity  experienced by the upstream system $\mathcal{U}$. If, however,  an insulating circuit needs to couple to multiple downstream subsystems (see Fig.~\ref{fig:schematic}.F), the retroactivity experienced by $\mathcal{I}$ may be relevant. In this case, the properties of those downstream interfaces could be analysed in the same way as we consider the retroactivity on $\mathcal{U}$.

\section{Results and Discussion}
\subsection{Dependence of retroactivity on the upstream subsystem}\label{sec:direct_binding}
In this section, we will consider three basic alternatives for the internal dynamics of the upstream subsystem $\mathcal{U}$, and explore the consequences for retroactivity. We will first illustrate the problem using a simple choice of $\mathcal{D}$ and the coupling between $\mathcal{U}$ and $\mathcal{D}$. We shall then seek to generalise the results.   
For our illustrative downstream subsystem $\mathcal{D}$, we consider the inter-conversion of molecular species $P$ and $C$.  We shall take a direct coupling between upstream and downstream subsystems via binding of $Z$ to $P$ to produce $C$. This setting is an extremely common motif for passing on a signal in biology, provided that the complex $C$ has properties that are distinct from those of $P$. For example, $Z$ could be a transcription factor that binds to a promoter $P$, triggering or suppressing translation~\cite{stock2000two}, or a receptor that recruits proteins to the cell membrane when active~\cite{Lemmon1996}. Similar motifs are widespread in nucleic acid nanotechnology~\cite{Qian2011,Green2014}.
\begin{enumerate}
\item\textbf{Fixed total concentration of $Z$:} In the simplest case, there is a fixed and finite pool of $Z$ molecules, $[Z_{\rm tot}] = [Z]+[C]$.  This description would approximate a setting in which a pool of $Z$ molecules are suddenly activated or released due to an external signal, and the overall system reaches a steady state response prior to the deactivation or recapture of $Z$ at the end of the signalling period. A fixed $[Z_{\rm tot}]$ during the signalling period is consistent with an {\it in vitro} setting in which there is no net turnover of components, or to an {\it in vivo} setting in which protein production and decay or dilution of molecules is slow compared to signal dynamics. In this case, we need only solve for the steady state of 
\begin{align}
Z+P&\xrightleftharpoons[k_{\text{off}}]{k_{\text{on}}}C
\end{align}
subject to $[Z_{\rm tot}] = [Z]+[C]$ and $[P_{\rm tot}] = [P]+[C]$. 
\item\textbf{Constant birth/death dynamics:} In this case, we assume that $Z$ molecules undergo a rapid  birth/death process in addition to binding to $P$. 
Specifically, we imagine that $Z$ molecules are produced and degraded at rate constants $k$ and $\delta$, respectively, as well as binding to $P$ to form the complex $C$. Thus the system
\begin{align}
\emptyset &\xrightleftharpoons[\delta]{k}Z\nonumber\\
Z+P&\xrightleftharpoons[k_{\text{off}}]{k_{\text{on}}}C
\end{align}
reaches steady state subject to the constraint $[P_{\rm tot}]=[P]+[C]$. Such a description would approximate an {\it in vivo} response to a variation of $k$ and $\delta$ on a time-scale slower than protein production and decay or dilution. This system was previously analysed in the stochastic setting by Ghaemi {\it et al.}~\cite{ghaemi}.
\begin{figure*}
\includegraphics[scale=0.615]{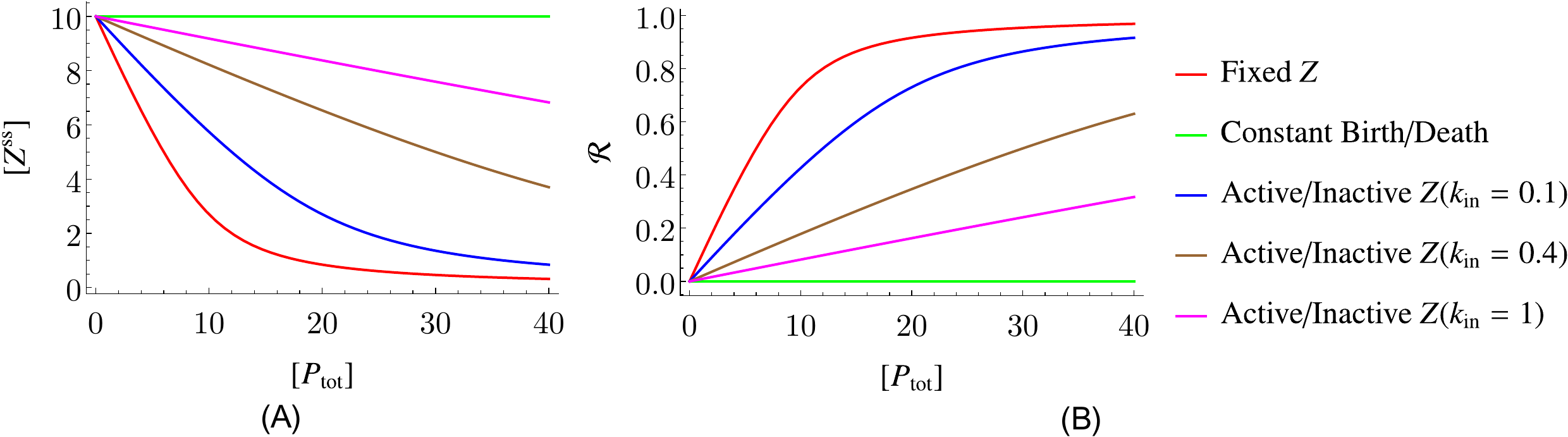}
\caption{Replenishing the pool of $Z$ from large reservoirs or by rapid production and decay of components can suppress retroactivity. To illustrate this, we plot the steady-state concentration of $Z$ and the retroactivity metric as a function of $[P_{\rm tot}]$ for three different upstream subsystems $\mathcal{U}$. Parameters of the system: $k = 10,\delta = k_{\text{on}} = k_{\text{off}} = 1, k_{\text{active}} = 0.1, [Z_{\text{tot}}]=\frac{k}{\delta}, [Z'_{\text{tot}}] = \frac{k_{\text{active}} + k_{\text{inactive}}}{k_{\text{active}}}[Z_{\text{tot}}].$
}\label{fig:variable_Z_dynamics}
\end{figure*}
\item\textbf{Active and inactive forms of $Z$:} Finally, we consider a setting in which $Z$ exists in both inactive ($Z_0$) and active ($Z$) forms, as well as in complex with $P$. We assume that $Z_0$ is incapable of forming a complex, and that the total population $[Z_{\rm tot}^\prime] = [Z_0] + [Z] + [C]$ is fixed. Such a setting would correspond to a situation similar to case (1), but when only a fraction of the the $Z$ molecules are activated or released in response to an external signal. In this case we solve 
\begin{align}
Z_0 &\xrightleftharpoons[k_{\text{in}}]{k_{\text{ac}}}Z\nonumber\\
Z+P &\xrightleftharpoons[k_{\text{off}}]{k_{\text{on}}}C
\end{align}
for the steady state subject to the constraints $[Z_{\rm tot}^\prime]= [Z]+[Z_0]+[C]$ and $[P_{\rm tot}] = [P]+[C]$. Here, $k_{\text{in}}$, and $k_{\text{ac}}$ are first-order rate constants,
\end{enumerate}
The steady-state concentration $[Z^{\rm ss}]$ is the output signal of $\mathcal{U}$; in the limit $[P_{\rm tot}] \rightarrow 0$ ($\mathcal{D}$ absent), the three alternatives for $\mathcal{U}$ all produce the same signal if 
\begin{align}
\frac{k}{\delta} = [Z_{\rm tot}] = \frac{k_{\rm ac}}{k_{\rm ac} + k_{\rm in}} [Z_{\rm tot}^\prime]. 
\end{align}
In Figure~\ref{fig:variable_Z_dynamics} we show $[Z^{\rm ss}]$ and the retroactivity metric $\mathcal{R}$ as we increase $[P_{\rm tot}]$, given equal $[Z^{\rm ss}]$ for $[P_{\rm tot}] \rightarrow 0$. Figure~\ref{fig:variable_Z_dynamics} demonstrates that retroactivity is highly sensitive to the internal details of $\mathcal{U}$. Clearly, the system with fixed $[Z_{\rm tot}]$ shows the strongest retroactivity; the system with constant birth and death of $Z$ shows no retroactivity; and the system with active and inactive forms of $Z$ interpolates between these two limits. Constant birth-death dynamics is analogous to having an infinite pool of $Z$ to draw upon (formally, $Z$ is coupled to a chemostat~\cite{novick1950description}). On average, a $Z$ molecule gets replenished every time it is consumed after binding to $P$. As a result, this system has zero retroactivity and $[Z^{\rm ss}] = \frac{k}{\delta}$ irrespective of $[P_{\rm tot}]$ -- this fact was previously noted in the stochastic setting by Ghaemi {\it et al.}~\cite{ghaemi}. The case with fixed $[Z_{\rm tot}]=[Z]+[C]$ has the highest retroactivity since there is nothing to replenish $Z$ once it binds to $P$. The setting with active and inactive forms of $Z$ implies a finite buffer upon which to draw; for low $[P_{\rm tot}]$, most of the sequestration of $Z$ can be compensated for by conversion of $Z_0$ into $Z$, but as $[P_{\rm tot}]$ grows, this buffer gets depleted. Consequently this third  case is moderately retroactive, interpolating between the regimes of fixed $[Z]+[C]$ and constant birth-death dynamics. In particular when $k_{\text{inactive}}\gg k_{\text{active}}$, this intermediate case approaches constant birth-death dynamics. Refer to Section~\ref{sec:analytics_Z_dynamics} in the Appendix for analytic expressions corresponding to these results.
Introducing birth-death dynamics for $Z$ is a general approach to buffering against the influence of downstream systems $\mathcal{D}$. In a wide range of steady-state contexts, this buffering eliminates retroactivity. In particular, $[Z]=k/\delta$ necessarily holds if the reaction network obeys detailed balance~\cite{aris1965prolegomena,aris1968prolegomena,feinberg1989necessary,gunawardena2003chemical,horn1972general}. However, a constant decay rate of the complex $C_1$ and a constant production of $P$ is sufficient to compromise this perfect buffering. Explicitly, the system
\begin{align}
&\emptyset \xrightleftharpoons[\delta]{k}Z,\,\,
&Z+P \xrightleftharpoons[\beta_2]{\beta_1}C_1 \nonumber\\
&C_1 \xrightarrow{\alpha_1} \emptyset,\,\, 
&\emptyset\xrightarrow{\gamma_1}P. 
\end{align}
has the following set of differential equations:
\begin{align}
\dot{[Z]} &= k - \delta[Z] -\beta_1[Z][P] + \beta_2[C_1] \nonumber\\
\dot{[C_1]} &= \beta_1[Z][P] - \beta_2[C_1] - \alpha_1 [C_1] \nonumber\\
\dot{[P]} &= -\beta_1[Z][P] + \beta_2[C_1] + \gamma_1. 
\end{align}
In steady-state, we have $[Z^{\rm ss}]=\frac{k-\gamma_1}{\delta}\neq \frac{k}{\delta}$ implying non-zero retroactivity. Increasing $\gamma_1$ increases retroactivity, as more $Z$ molecules are consumed by the downstream system and never released.
Even in the case of perfect buffering, implementing a system that produces and degrades components on a time scale that is fast compared to signal variation would be extremely expensive. Turning over a single protein molecule, for example, costs a cell on the order of thousands of ATP fuel molecules~\cite{bender2014introduction}. The alternative of having a very large but fixed pool of molecules from which to create $Z$ is also costly; energy (and in the cell, space) needs to be devoted to these molecules, and in vivo the large population would need to be maintained against a background dilution/decay, which is more costly than maintaining a small population. Similar arguments apply to maintaining a large pool of $Z$ that bind to $P$ only weakly, or more complex  $\mathcal{U}$ subsystems that replenish $Z$. 
In summary, although the use of a large or bottomless supply of $Z$ can suppress retroactivity, the inherent cost of this strategy is related to the cost of producing a large number of molecules, which is generally high. In  Section~\ref{subsec:push_pull}, we consider the alternative of using an insulating push-pull motif, in which retroactivity is suppressed without excessive production of the signalling species. Instead, an energy-consuming circuit involving catalytic molecular modification is used to reduce sequestration of $Z$. This modification consumes chemical fuel molecules such as ATP, which are far less costly than, for example, entire proteins. It is because of the relatively low cost of post-translational modification that we focus on the push-pull motif, rather than alternative insulators incorporating protein production and degradation \cite{ninfa}.

\subsection{The relationship between retroactivity and fuel consumption for an insulating push-pull motif}\label{subsec:push_pull}
Having identified the costs of suppressing retroactivity through the design of $\mathcal{U}$, we now turn to the alternative approach of using insulating motifs as shown in Fig.~\ref{fig:schematic}.E.ii. We reiterate the question first raised by \cite{sontag_barton}: is increased consumption of fuel necessary for suppression of retroactivity in a push-pull network? To approach this question, we will  consider an insulating push-pull motif for an upstream network $\mathcal{U}$ with a fixed total amount of $Z$. This was the simplest $\mathcal{U}$ considered in Section~\ref{sec:direct_binding}. This choice is motivated not only by simplicity, but also because it is the most challenging context for an insulator (underlying retroactive effects are strongest) and because the alternative choices of $\mathcal{U}$ are associated with their own additional resource costs. Our system is therefore defined by the set of reactions below, following Ref.~\cite{sontag_barton}:
\begin{align}
Z+X &\xrightleftharpoons[\beta_2]{\beta_1}C_1\xrightarrow{k_1} X^* + Z, \nonumber\\
Y+X^*&\xrightleftharpoons[\alpha_2]{\alpha_1}C_2\xrightarrow{k_2} X + Y, \nonumber\\
X^*+p&\xrightleftharpoons[k_{\text{off}}]{k_{\text{on}}}C.
\label{eq:push-pull}
\end{align}
Here we have assigned mass-action rate constants to each step, and assumed that the role of the molecular fuel molecules ATP, ADP and P$_{\rm i}$ can be implicitly absorbed into rate constants, as is common. We have also assumed that the free energy of ATP breakdown, $\Delta G_{\rm ATP}$, is sufficiently large that dephosphorylation via $Z$ and phosphorylation via $Y$ are never observed. \\
An explicit representation of the intermediate catalyst-substrate complexes $C_1$ and $C_2$ allows for a quantification of sequestration of $Z$ by the push-pull insulator. Again, we consider the fractional reduction in free $Z$ due to the introduction of the downstream system:
\begin{equation}
\mathcal{R} = \left| 1-\frac{ [Z^{\rm ss}]}{[Z^{\rm ss}_{{\mathcal{D,I} \rightarrow \emptyset}}]}\right| = \frac{[C^{\rm ss}_1]}{[Z_{\rm tot}]}.
\end{equation}
The system illustrated in Eq.~\ref{eq:push-pull} turns over a single molecule of ATP per phosphorylation/dephosphorylation cycle, in which a molecule of $X$ is first activated by $Z$ then deactivated by $Y$. Thus the fuel consumption rate per unit volume is given by the net flux $\Psi$ of $X$ molecules around this cycle. The overall power per unit volume is given by $w=\Psi\Delta G_{\rm ATP}$, where $\Delta G_{\rm ATP}$ is the free energy released by the breakdown of a single ATP molecule. From inspection, 
\begin{equation}
\Psi = k_1 [C_1^{\rm ss}], \hspace{5mm} w =  k_1 [C_1^{\rm ss}]\Delta G_{\rm ATP}.
\end{equation}
Strictly speaking, $\Delta G_{\rm ATP}$ should be infinite since we have approximated the catalytic reactions as microscopically irreversible, as in~\cite{sontag_barton,ninfa,stadtman1977superiority,goldbeter1981amplified}. In practice, $\Delta G_{\rm ATP}$ is assumed to have a large, fixed, negative value. The power consumption will then be essentially determined by the flux $\Psi$ through the cycle. As a consequence, we see that both retroactivity and fuel consumption grow proportionally to $[C_1^{\rm ss}]$. In intuitive terms both fuel turnover through phosphorylation, and the retroactivity, increase if $Z$ is frequently bound to $X$ to form the enzyme-substrate complex $C_1$. This observation raises questions about the conclusion that increased energy consumption is necessary to {\em suppress} retroactivity from Ref.~\cite{sontag_barton}.
To explore this idea further, let us consider whether an insulating system can in general be tuned to reduce both retroactivity and fuel consumption whilst maintaining its signal-transducing function. Specifically, let us ask whether combinations of system parameters can be changed so that the input-output relation $[C^{\rm ss}]([Z_{\rm tot}])$ is approximately preserved, but both $w$ and $\mathcal{R}$ are reduced. In Refs.~\cite{ninfa,sontag_barton,shah2017signaling,legewie2008recurrent,dekel2005optimality}, optimisation of insulating circuits was only performed over the total concentrations of species. Natural evolution or deliberate design will, however, allow for moderation of at least some of the chemical rate constants, and we will focus on these parameters.
If it were possible to take the catalytic rate constants $k_1,k_2 \rightarrow \infty$, we would obtain ${[C^{\rm ss}_1]}\rightarrow 0$ and $\mathcal{R}\rightarrow 0$, and retroactivity could be completely eliminated, regardless of the fuel consumption rate. However, these catalytic rate constants encode complex chemistry that is likely difficult to accelerate. We shall therefore take $k_1,k_2$ as fixed and instead focus on $\alpha_1,\beta_1$, the rate constants of enzymatic binding. Whilst there is a diffusion-based limit to how far $\alpha_1,\beta_1$ can be {\em increased}~\cite{tenWolde2016}, we assume that it is always possible to reduce the speed with which a given catalyst binds to its substrate. Alternatively, we could have focussed on  $\alpha_2,\beta_2$ (the unbinding rate constants), assuming that it is always possible to increase them. In both cases, which yield similar results, we essentially assume that it is possible to reduce the catalyst-substrate binding affinity, either through design of synthetic systems or via evolution of natural systems. {For example, this reduced affinity could arise from replacing a hydrophobic residue by a hydrophilic one in a protein-protein interaction or creating a mismatch in DNA-DNA interaction.}
\begin{figure*}
\includegraphics[scale=0.46]{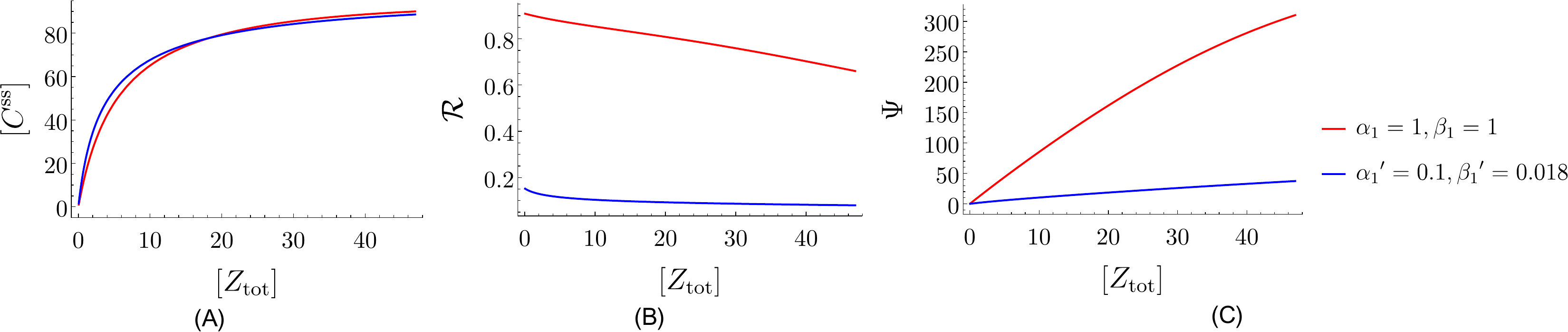}
\caption{An example of reducing both fuel consumption and retroactivity whilst maintaining an approximately fixed input-output relation, even with high initial retroactivity. (A) Input-output relation for two systems with distinct $\alpha_1$, $\beta_1$, but otherwise identical parameters. The second curve is obtained by setting $\beta_1^\prime = 0.018\beta_1$, and adjusting $\alpha_1^\prime$ to maximise the similarity between curves. (B) Retroactivity $\mathcal{R}$ for the system in (A), illustrating substantially lower retroactivity for $\alpha_1^\prime, \beta_1^\prime < \alpha_1, \beta_1$. (C) Flux $\Psi$, 
which is much reduced for $\alpha_1^\prime, \beta_1^\prime < \alpha_1, \beta_1$. Other parameters: $[X_{\text{tot}}]=200,[Y_{\text{tot}}]=100,[P_{\text{tot}}]=100,\alpha_2=\beta_2=k_1=k_2=k_{\text{on}}=k_{\text{off}}=10$.}\label{fig:retroactivity_irreversibility}
\end{figure*} 
We first consider the low retroactivity limit, when $[C_1^{\rm ss}]$ and $[C_2^{\rm ss}]$ are both small. In this case,
\begin{equation}\label{eq:irreversible_output}
[C^{\rm ss}]\approx\frac{f(r)\pm\sqrt{f^2(r) - 4r^2{k^2_{\text{on}}}[X_{\text{tot}}][P_{\text{tot}}]}}{2rk_{\text{on}}}
\end{equation}
where 
\begin{align}
r = \frac{\beta_1}{\alpha_1} \frac{k_1[Z_{\text{tot}}]}{k_2[Y_{\text{tot}}]}\frac{k_2 + \alpha_2}{\beta_2+k_1},
\end{align}
and 
\begin{align}
f(r)=k_{\text{off}} + r(k_{\text{on}}([P_{\text{tot}}] + [X_{\text{tot}}]) + k_{\text{off}}).\end{align}
In this limit, $[C^{\rm ss}]$ is a function of $\beta_1/\alpha_1$, rather than $\alpha_1$ and $\beta_1$ independently, if all other parameters are fixed. Moreover, 
\begin{eqnarray}\label{eq:irreversible_retroactivity_flux}
\mathcal{R},\Psi\propto\beta_1\text{ at fixed }\frac{\beta_1}{\alpha_1}
\end{eqnarray}
(see Section~\ref{low_retroactivity} in the Appendix for a derivation of these facts). If we then reduce $\beta_1$ and $\alpha_1$ by the same factor $\phi$ whilst keeping all other parameters fixed, the input-output relation $[C^{\rm ss}]([Z_{\rm tot}])$ is unchanged whilst $\mathcal{R}$ and $w$ are both reduced by $\phi$. In principle, this simultaneous reduction of retroactivity and fuel consumption at fixed input-output relation can proceed arbitrarily far.
The above observation forms the intuition behind the main claim of this paper. Fundamentally, the push-pull network responds to a competition between $Z$ and $Y$. We can therefore reduce the strength with which both $Z$ and $Y$ couple to $X$, whilst maintaining the same steady-state output. Reducing the coupling to $X$ serves to minimise both retroactivity and energy consumption. 
If $[C_1^{\rm ss}]$ and $[C_2^{\rm ss}]$ are not both small, the input-output relationship is not a function of $\frac{\beta_1}{\alpha_1}$ only. It is therefore no longer possible to  reproduce input-output relations exactly as outlined above. However, we can instead consider reducing $\beta_1 \rightarrow \beta_1^\prime$, and identifying the corresponding change $\alpha_1\rightarrow \alpha_1^\prime$ that reproduces the original input-output curve as closely as possible. Specifically, we identify the new $\alpha_1^\prime$ as the value that minimizes the following measure of the difference between input-output relations
\begin{align}\label{eq:input_output_parameters}
\Int_{l}^{u}\mid &[C^{\rm ss}]([Z_{\text{tot}}],\alpha_1,\beta_1)-
\nonumber \\ 
& [C^{\rm ss}]([Z_{\text{tot}}],\alpha_1^\prime,\beta_1^\prime)\mid\,{\rm d}[Z_{\text{tot}}],
\end{align}
where $l$ and $u$ are such that $C^{\rm ss}(l,\alpha_1,\beta_1)\approx 0.01[P_{\text{tot}}]$ and $C^{\rm ss}(u,\alpha_1,\beta_1)\approx 0.9[P_{\text{tot}}]$.
\begin{figure*}[!]
  \subfigure{\includegraphics[scale=0.5]{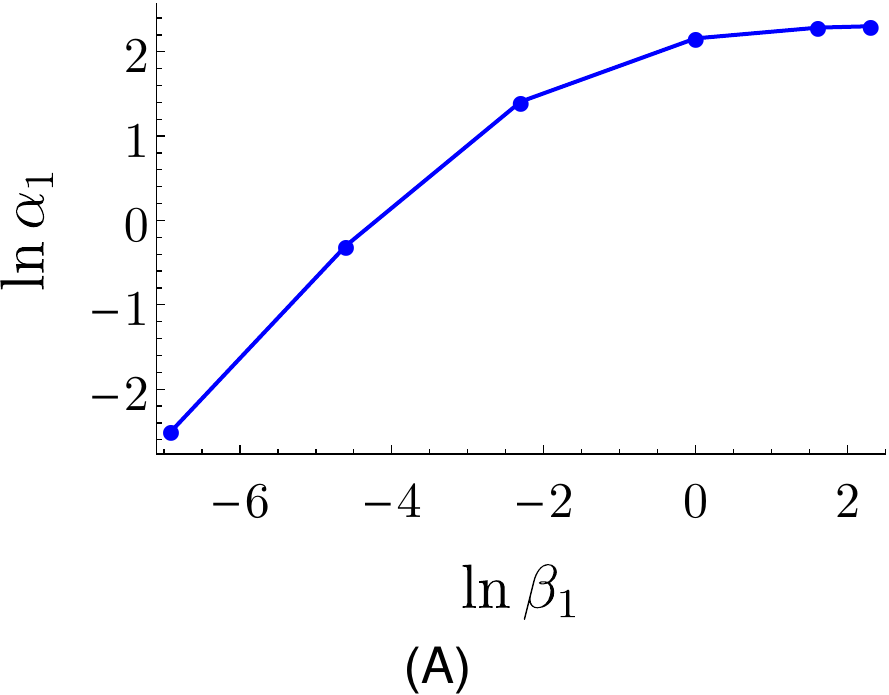}}\hspace{8mm}
  \subfigure{\includegraphics[scale=0.5]{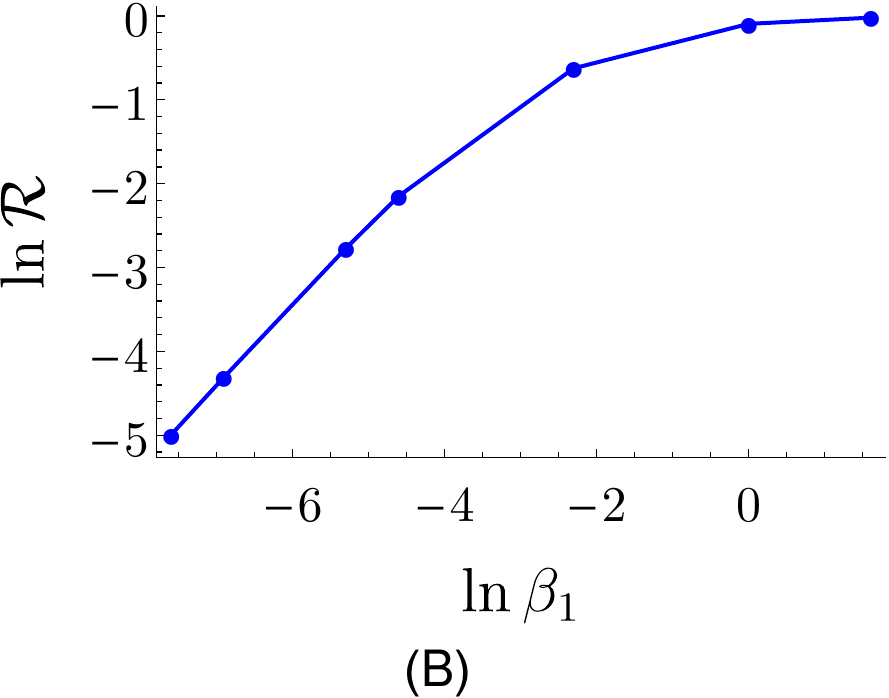}}\hspace{8mm}
  \subfigure{\includegraphics[scale=0.5]{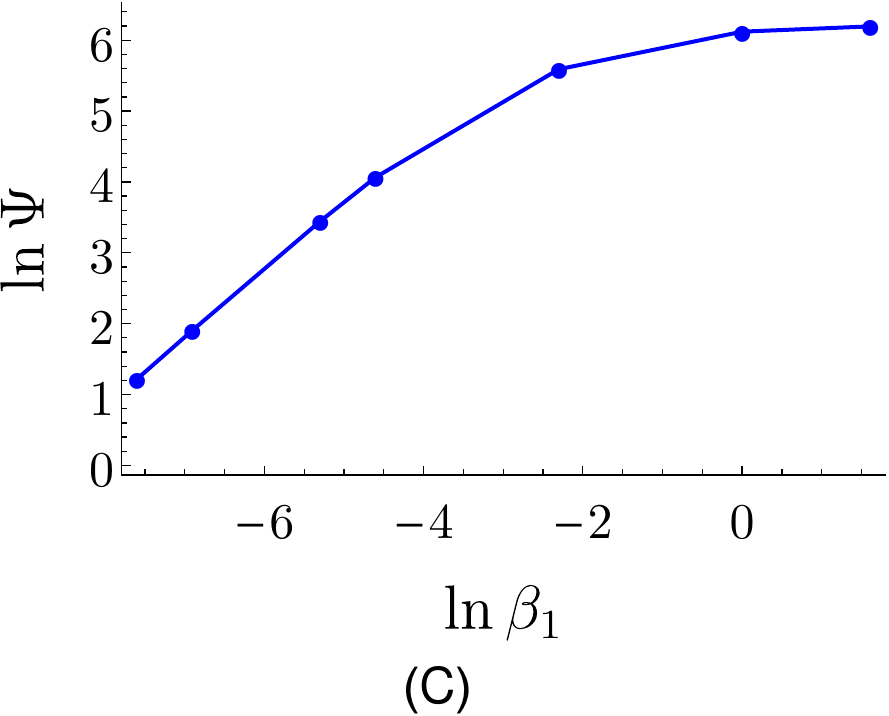}}
  \caption{In the limit of low $[C_1^{\rm ss}]$ and $[C_2^{\rm ss}]$ or equivalently low $\alpha_1$ and $\beta_1$, the steady-state concentration of the output $[C^{\rm ss}]$ depends on the ratio $\frac{\beta_1}{\alpha_1}$ and not $\alpha_1$ and $\beta_1$ individually. (A) Scaling of $\alpha_1$ and $\beta_1$ as they are simultaneously adjusted to retain a given input-output curve. (B) Scaling of retroactivity $\mathcal{R}$ with $\beta_1$ as this operation is performed. (C) Scaling of net flux $\Psi$ with $\beta_1$ as this operation is performed. Both the flux and retroactivity decrease as $\beta_1$ is decreased, becoming 
proportional to $\beta_1$ at fixed ratio $\frac{\beta_1}{\alpha_1}$ in the low retroactivity limit. Constant parameters for the network: $[X_{\text{tot}}]=400,[Y_{\text{tot}}]=150,[Z_{\text{tot}}]=50,[P_{\text{tot}}]=100,\alpha_1=10,\alpha_2=15,\beta_1=10,\beta_2=15,k_1=k_2=k_{\text{on}}=k_{\text{off}}=10$.}\label{fig:ratio_alpha_beta}
\end{figure*}
Even when retroactivity (and hence $[C_1^{\rm ss}]$) is high for the original parameters $\alpha_1,\beta_1$, it is frequently possible to approximate the input-output relation with reduced $\alpha_1^\prime,\beta_1^\prime$. Consider, for example, the input-output curves in Fig.\ref{fig:retroactivity_irreversibility}, in which $\mathcal{R}\sim 0.8$. As expected, the reduced $\alpha_1^\prime,\beta_1^\prime$ give substantially lower retroactivity and fuel consumption. Further examples are provided in Section~\ref{subsec:example_finite_free_energy} of the Appendix.
 
The above process can be iterated, producing ever lower retroactivity and energy consumption by a continuing reduction in coupling strength between $Z$ and the insulator. We illustrate the process in Fig.~\ref{fig:ratio_alpha_beta}. Eventually, the limit of small $[C_1^{\rm ss}]$ and $[C_2^{\rm ss}]$ is reached and $\alpha_1$ and $\beta_1$ decrease in proportion, with $\mathcal{R}$ and $\Psi$ also scaling proportionally. 

\subsection{Incorporating microscopic reversibility for an insulating push-pull motif}\label{subsec:finite_free_energy}
\begin{figure*}[t]
  \subfigure{\includegraphics[scale=0.7]{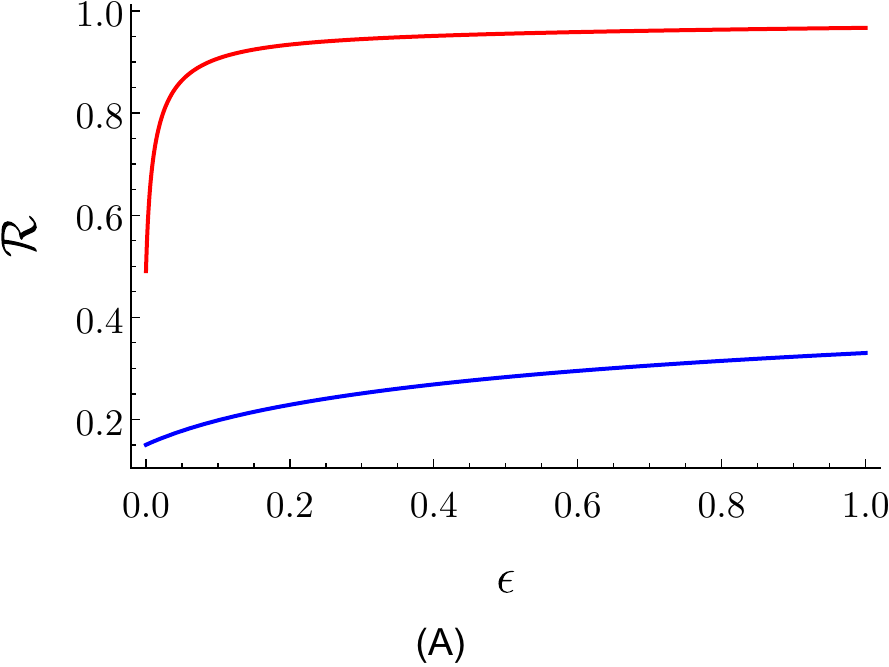}}\hspace{12mm}
  \subfigure{\includegraphics[scale=0.7]{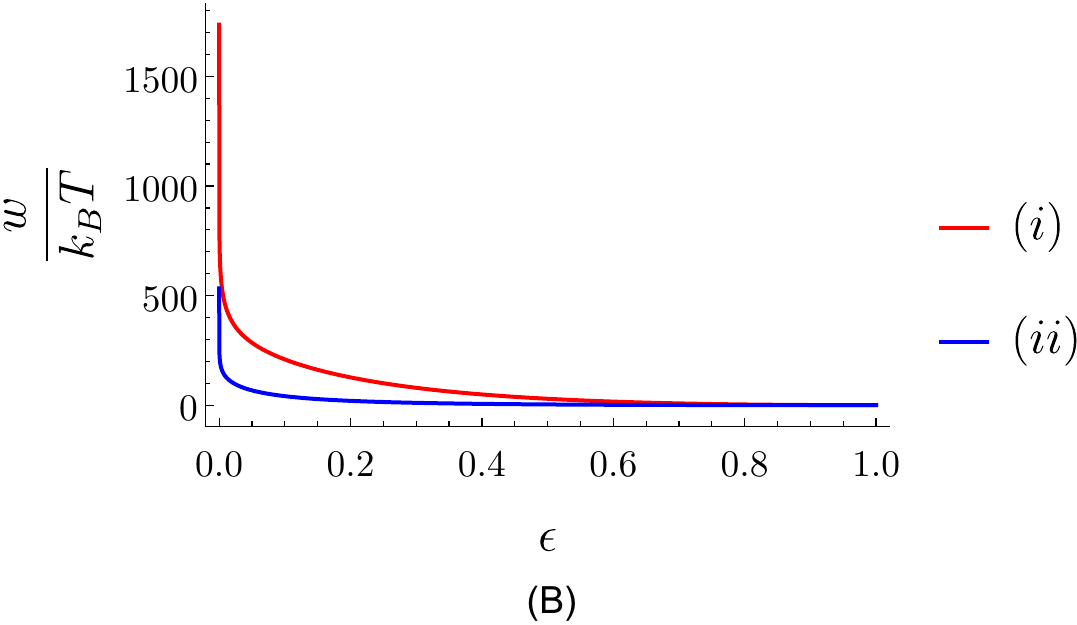}}
  \caption{A) Inclusion of microscopically reversible reactions can increase the retroactivity of the motif due to rebinding of products to catalysts; $\mathcal{R}$ increases with $\epsilon$ for a specific system while all other parameters are fixed. B) The rate of energy consumption decreases with increase in $\epsilon$. However, this is not primarily due to a decrease in flux, but rather due to the decrease in free energy of ATP molecules. Parameters of the network: $ i)\hspace{2 mm}[X_{\text{tot}}]=300,[Y_{\text{tot}}]=50,=[Z_{\text{tot}}]=100,[P_{\text{tot}}]=100,\alpha_1=\alpha_2=\beta_1=\beta_2=k_1=k_2=1,k_{\text{on}}=k_{\text{off}}=10. (ii)\hspace{2 mm} [X_{\text{tot}}]=[Y_{\text{tot}}]=[Z_{\text{tot}}]=[P_{\text{tot}}]=100,\alpha_1=\alpha_2=\beta_1=\beta_2=0.1,k_1=k_2=1,k_{\text{on}}=k_{\text{off}}=10.$}\label{fig:increase_retroactivity}
\end{figure*}

\begin{figure*}[t]
\centering
\includegraphics[scale=0.46]{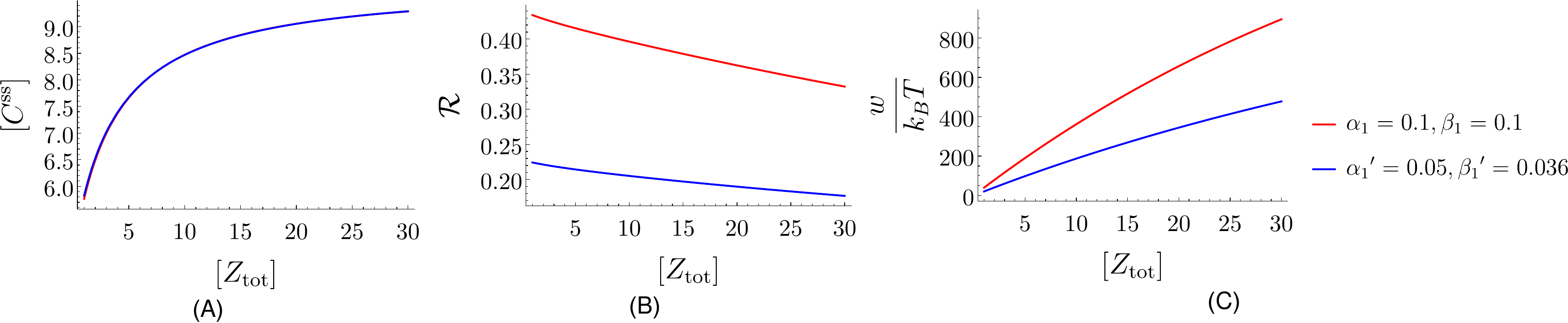}
\caption{Simultaneous reduction in retroactivity and energy consumption whilst approximately maintaining the input-output relation at a fixed and finite free energy stored per fuel molecule. (A) Two different sets of binding rates $\alpha_1,\beta_1$ and $\alpha^\prime_1,\beta_1^\prime$ that give a similar input-output relation with all other parameters fixed ($\alpha^\prime_1$ is chosen by minimising Eq.~\ref{eq:input_output_parameters} for given $\alpha_1,\beta_1, \beta_1^\prime$). (B) and (C) show retroactivity $\mathcal{R}$ and power $w$ for the two cases. Other Parameters used $[X_{\text{tot}}]=[Y_{\text{tot}}]=100,[P_{\text{tot}}]=10,\alpha_1=\beta_1=0.1,\alpha_2=\beta_2=1,k_1=k_2=k_{\text{on}}=k_{\text{off}}=10,\epsilon=0.01$.}\label{fig:retroactivity_energy_reversibility}
\end{figure*}

In principle, all chemical reactions are microscopically reversible~\cite{Lewis01031925,Ouldridge_review_2017}. This fact is often ignored when studying physiological ATP-driven systems, as it was by Barton and Sontag~\cite{sontag_barton}, since the high free energy of ATP hydrolysis~\cite{zubay1998biochemistry} can render reverse reactions irrelevant to the eventual steady state. Such an approximation is typically reasonable for push-pull motifs driven by free energies substantially in excess of $4k_BT$~\cite{Govern09122014}, as we confirm in our case in Section~\ref{sec:4kT_criterion} of the Appendix. Nonetheless, a full understanding of the resource requirements of insulators requires explicit treatment of reverse reactions, since their contribution depends directly on the free energy consumed per cycle. Moreover, in developing synthetic systems, a chemical fuel with a free energy as high as physiological ATP may be unavailable or undesirable. We therefore explicitly incorporate microscopic reversibility into our discussion in this session. We introduce microscopically reversible reactions in the simplest possible way, still assuming a single long-lived catalyst/substrate complex:
\begin{align}
Z+X &\xrightleftharpoons[\beta_2]{\beta_1}C_1\xrightleftharpoons[\frac{\epsilon k_1\beta_1}{\beta_2}]{k_1} X^* + Z, \nonumber\\
Y+X^*&\xrightleftharpoons[\alpha_2]{\alpha_1}C_2\xrightleftharpoons[\frac{\epsilon k_2\alpha_1}{\alpha_2}]{k_2} X + Y, \nonumber\\
X^*+p&\xrightleftharpoons[k_{\text{off}}]{k_{\text{on}}}C.
\end{align}
Here, $\epsilon$ is the parameter that modulates the distance of the system from equilibrium; $\epsilon=0$ corresponds to a completely irreversible network, with infinite driving, while $\epsilon=1$ corresponds to an equilibrium network. Equivalently, the free energy of the molecular fuel consumed in a single cycle is  
%\begin{eqnarray}\label{eq:free_energy_reversible}
$
\Delta G_{\rm ATP}=2k_BT \ln\epsilon
$.
%\end{eqnarray}
\begin{figure*}[t]
\subfigure{\includegraphics[scale=0.7]{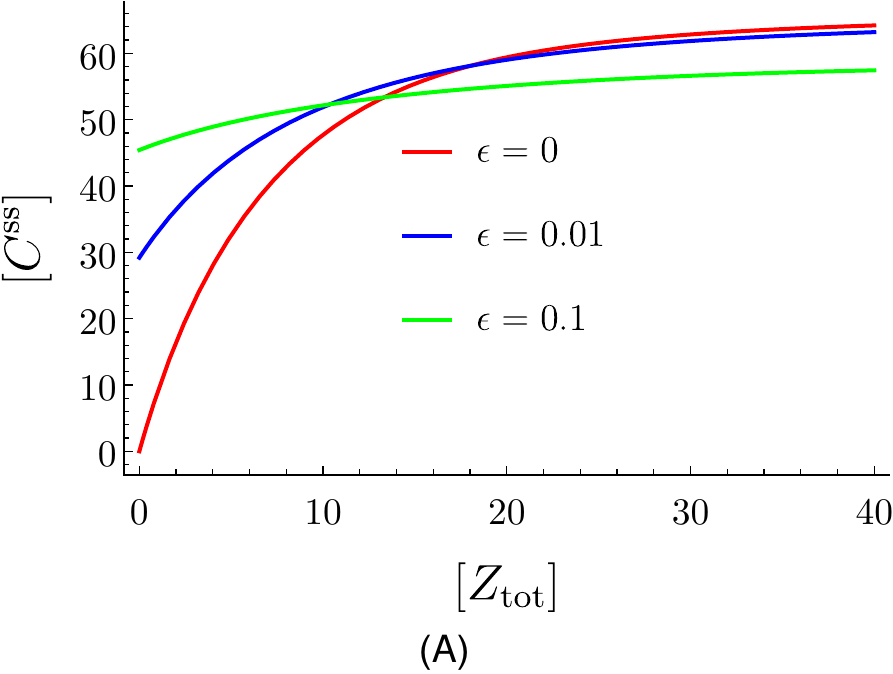}}\hspace{26mm}
\subfigure{\includegraphics[scale=0.7]{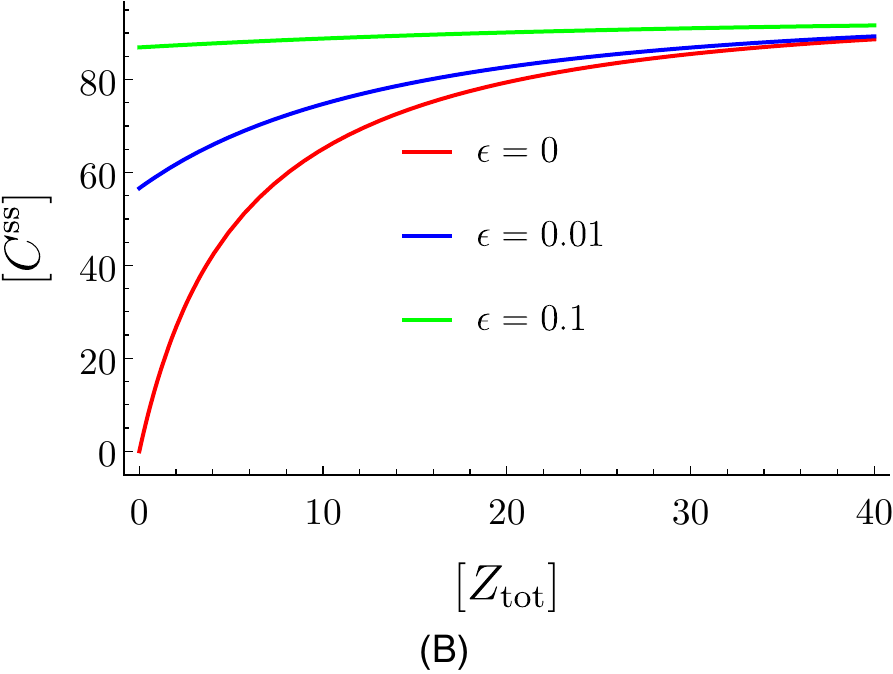}}
\caption{Increasing microscopic reversibility limits the ability of the network to produce a wide range of output. Parameters used for the network: $\text{A})\hspace{1mm} [X_{\text{tot}}]=100,[Y_{\text{tot}}]=100,[P_{\text{tot}}]=100,\alpha_1=\beta_1=\alpha_2 =\beta_2 = 0.1,k_1=k_2=1,k_{\text{on}}=k_{\text{off}}=10.\hspace{1mm}\text{B})\hspace{1mm}[X_{\text{tot}}]=200,[Y_{\text{tot}}]=100,[P_{\text{tot}}]=100,\alpha_1=\beta_1=0.1,\alpha_2=\beta_2=k_1=k_2=1,k_{\text{on}}=k_{\text{off}}=10.$}
\label{fig:output_reversibility}
\end{figure*}
Fundamentally, imposing a finite free energy per fuel molecule through microscopic reversibility limits the overall range of the input-output function $[C^{\rm ss}]([Z_{\rm tot}])$. Intuitively, if catalysts function in both directions, $[X]>0$ even if $[Z_{\rm tot}]/[Y_{\rm tot}] \rightarrow \infty$. Similarly, $[X^*] > 0
$ even if $[Z_{\rm tot}]/[Y_{\rm tot}] \rightarrow 0$. Indeed, 
\begin{align}
\epsilon \leq \frac{[X^*]}{[X]} \leq \frac{1}{\epsilon},
\end{align} 
implying a reduced dynamic range of the insulator and hence a weaker propagation of the signal from $Z$ to the output. We illustrate this intuition for a specific system in Figure~\ref{fig:output_reversibility}. The overall range of the input-output function drops as $\epsilon$ increases from 0 towards 1. 
Although potentially problematic for signal propagation, the reduction in range of the input-output function $[C^{\rm ss}]([Z_{\rm tot}])$ is not an inherently retroactive effect. It is not a direct consequence, nor a cause, of sequestration of $Z$ by the downstream subsystem. We now explicitly consider the effect of microscopic reversibility on retroactivity. We show  that decreasing the free energy consumed per fuel molecule can increase retroactivity due to rebinding of products to catalysts. However, the essential arguments of Section~\ref{subsec:push_pull} remain valid; it is still possible to simultaneously reduce free-energy consumption and retroactivity through weak coupling of $Z$ and $Y$ to $X$.
For the push-pull motif that explicitly incorporates the microscopically reversible reactions,
\begin{align}
\Psi=k_1[C_1^{\rm ss}] - \frac{\epsilon k_1\beta_1}{\beta_2}[X^{*\text{ss}}][Z^\text{ss}],
\end{align}
and 
\begin{align}
w=2k_BT\ln\epsilon\left(k_1[C_1^{\rm ss}] - \frac{\epsilon k_1\beta_1}{\beta_2}[X^{*\text{ss}}][Z^\text{ss}]\right),
\end{align}
whilst the retroactivity remains $\mathcal{R} = [C_1^{\rm ss}] /[Z_{\rm tot}]$. Away from the microscopically irreversible limit of $\epsilon \rightarrow 0$, retroactivity and energy consumption are less directly related than  in Section~\ref{subsec:push_pull}. Indeed, if we simply keep all other parameters fixed whilst varying $\epsilon$, it is possible to simultaneously increase retroactivity and decrease overall power consumption (or vice-versa). In particular, both $\Psi$ and $w$ tend to decrease as $\epsilon \rightarrow 1$, but $\mathcal{R}$ can be enhanced as both $X$ and $X^*$ can bind to $Z$ to produce $C_1$. A specific 
example is given in Fig.~\ref{fig:increase_retroactivity}.  
The above observation, however, does not imply that reduction of retroactivity necessarily requires high rates of free energy consumption. In particular, for a push-pull motif of fixed $\epsilon$, we can play essentially the same trick as before: reduce the strength of coupling to the push-pull by decreasing both $\alpha_1$ and $\beta_1$ in such a way that approximately maintains the input-output relation $[C^{\rm ss}]([Z_{\rm tot}])$. In fact, just as in the completely irreversible case, one can show that in the limit of low $[C_1^{\rm ss}]$ and $[C_2^{\rm ss}]$, the steady-state output is a function of $\frac{\beta_1}{\alpha_1}$ but not $\alpha_1$ and $\beta_1$ separately. Moreover,
\begin{eqnarray}\label{eq:reversible_retractivity_power}
\mathcal{R},w\propto\beta_1\text{ at fixed }\frac{\beta_1}{\alpha_1}
\end{eqnarray}
still holds (see Section~\ref{low_retroactivity} of the Appendix). In the limit of low retroactivity, one can therefore decrease both $\alpha_1$ and $\beta_1$ in proportion to give the same input/output curve at reduced retroactivity and energy consumption, as before.
For higher $[C_1^{\rm ss}]$ and $[C_2^{\rm ss}]$, just as in Section~\ref{subsec:push_pull}, it is not possible to obtain the same input-output curve by varying $\beta_1$ at fixed $\frac{\beta_1}{\alpha_1}$. However, we again observe that in most cases a good fit to the input-output relation can be obtained by reducing $\beta_1$ and $\alpha_1$ in such a way as to minimize Eq.~\ref{eq:input_output_parameters}, even when $\mathcal{R}$ is appreciable, in the process reducing both $\mathcal{R}$ and $w$. We demonstrate this behaviour for a specific system in Fig.~\ref{fig:retroactivity_energy_reversibility}; other examples are given in Section~\ref{subsec:example_finite_free_energy} of the Appendix.
\subsection{Arbitrarily weak coupling to an insulator causes vulnerability to cross-talk}\label{sec:leak_reactions}
Biochemical signalling pathways do not exist in isolation; in both natural and complex synthetic systems multiple information transmission pathways based on similar reactions must co-exist~\cite{friedlander2016intrinsic,Qian2011,Swain2002,Ouldridge_proofreading_2014}. Transferring information to only the desired downstream recipients is a challenge in specificity; the possibility of unintended interference would compromise information transduction. In this section we demonstrate how cross-talk limits the degree to which weak  insulator-coupling allows effective signalling with low fuel consumption and low retroactivity. To do so we consider the system in Eq.~\ref{eq:leak}, with an additional upstream molecule $Z^\prime$ that couples to $X$ through an accidental leak reaction. As a result, we get the following network:
\begin{align}
Z^\prime+X &\xrightleftharpoons[\gamma_2]{\gamma_1}C_4\xrightarrow{k_3} X^* + Z^\prime, \nonumber\\
Z+X &\xrightleftharpoons[\beta_2]{\beta_1}C_1\xrightarrow{k_1} X^* + Z, \nonumber\\
Y+X^*&\xrightleftharpoons[\alpha_2]{\alpha_1}C_2\xrightarrow{k_2} X + Y, \nonumber\\
X^*+p&\xrightleftharpoons[k_{\text{off}}]{k_{\text{on}}}C.
\label{eq:leak}
\end{align}
\begin{figure*}[t]
  \subfigure{\includegraphics[scale=0.7]{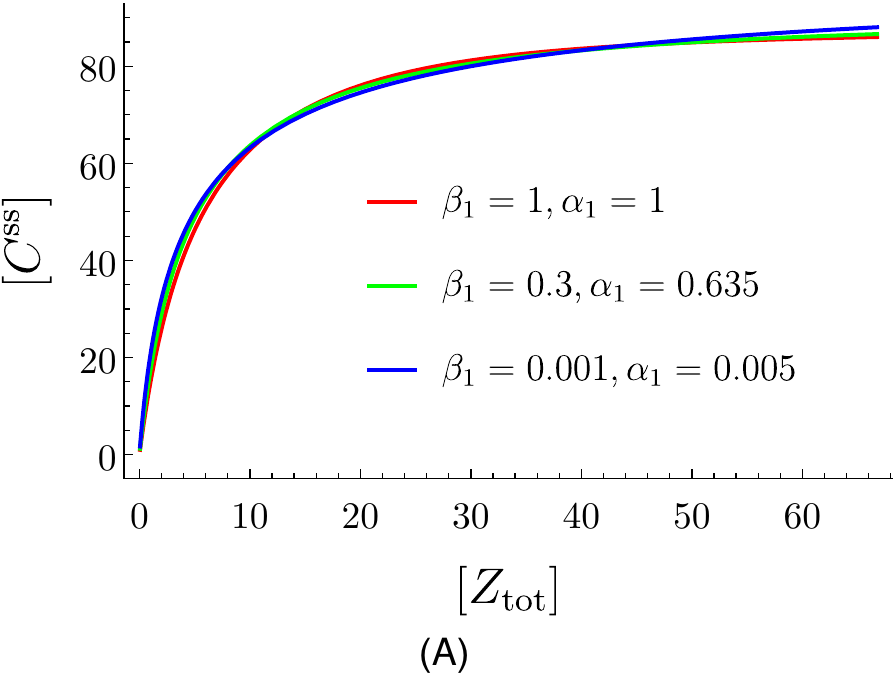}}\hspace{23mm}
  \subfigure{\includegraphics[scale=0.7]{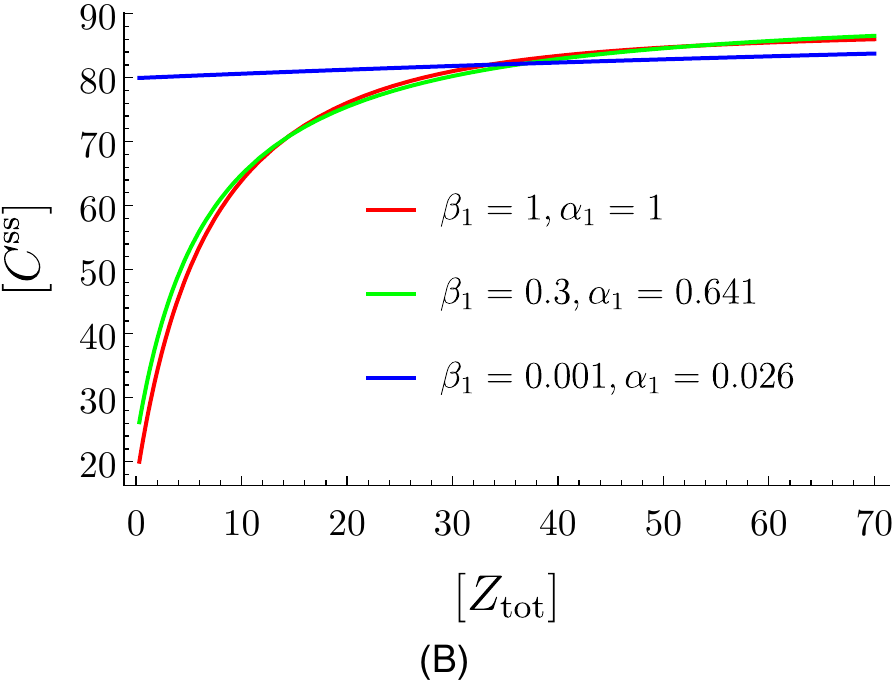}}
  \caption{Evidence that accidental leak reactions limit the degree to which coupling between $\mathcal{U}$ and $\mathcal{I}$ can be reduced whilst maintaining a functioning network. (A) In the absence of a leak reaction, one can decrease $\alpha_1$ and $\beta_1$ successively using exactly the procedure outlined in Section~\ref{subsec:push_pull}, maintaining an approximately constant input-output curve. (B) In the presence of a leak reaction caused by a signal molecule $Z^\prime$, one is able to reduce $\alpha_1$ and $\beta_1$ to match a given input-output curve whilst the coupling of $Z^\prime$ remains relatively weak. However, eventually $Z^\prime$ becomes dominant and one cannot find suitable parameters $\alpha_1$ and $\beta_1$ to match a given input-output curve well. Parameters used for both networks: $[X_{\text{tot}}]=150,[Y_{\text{tot}}]=100,[Z'_{\text{tot}}]=100,[P_{\text{tot}}]=100,\alpha_1=\beta_1=1,\alpha_2=\beta_2=10,\gamma_1=0.01,\gamma_2=k_1=k_2=k_{\text{on}}=k_{\text{off}}=10,k_3=1$.}
 \label{fig:gamma_coupling} 
\end{figure*}
Conceptually, $Z^\prime$ represents the combined effect of many alternative catalysts that could cause accidental activation of $X$ via a leak reaction. It will therefore be challenging to reduce $\gamma_1$ arbitrarily far, either by evolution or design, whilst retaining the functions of these alternative catalysts within their intended pathways. In Fig.~\ref{fig:gamma_coupling}, we repeat the protocol of progressively weakening $\alpha_1$ and $\beta_1$ while attempting to preserve the input-output curve $[C^{\rm ss}]([Z_{\rm tot}])$ as outlined in Section~\ref{subsec:push_pull}, comparing a system with $\gamma_1=0$ to a system with a fixed and finite $\gamma_1$.
Crucially, we now observe that as the coupling between $Z$ and $X$ becomes weaker, $Z^\prime$ starts to dominate the insulator's behaviour. When the coupling between $Z^\prime$ and $X$ exceeds the coupling between $Z$ and $X$, we are no longer able to reduce $\alpha_1$ and $\beta_1$ to give a close match to the original curve, since $[C^{\rm ss}]$ responds primarily to $Z^\prime$ rather than $Z$. The range of the input-output function $[C^{\rm ss}]([Z_{\rm tot}])$ is consequently compromised, and signal propagation becomes ineffective.
The strength of leak reactions or cross-talk thus determines the degree to which effective signalling can be maintained despite weak coupling between $\mathcal{U}$ and the insulator $\mathcal{I}$. The system $\mathcal{U}$ must couple more strongly than cross-talk reactions, and consequently  fuel turnover and retroactivity cannot be suppressed arbitrarily far whilst retaining a functioning network. A similar consideration shows that the rates of spontaneous phosphorylation and dephosphorylation also limit the degree to which $Z$ can couple weakly to the insulator and remain effective. It is important to note, however, the logical distinction between the observation that some degree of fuel turnover and retroactivity are required for effective signal propagation, and the suggestion that increased fuel consumption is required to suppress retroactivity.

\section{Conclusions}\label{sec:conclusion}

We have considered the suppression of retroactivity in molecular signal transduction systems by both the design of the upstream subsystem $\mathcal{U}$, and by incorporating an insulator $\mathcal{I}$ between the $\mathcal{U}$ and the downstream subsystem $\mathcal{D}$. Using the fractional reduction in the concentration of the output of $\mathcal{U}$ due to the presence of $\mathcal{D}/\mathcal{I}$ as a metric for retroactivity~\cite{ninfa,shah2017signaling}, we find that retroactivity is strongly dependent on the design of $\mathcal{U}$, and that insulators can suppress retroactivity at low levels of fuel consumption. 

In particular, if $\mathcal{U}$ consists of a single species $Z$ undergoing production and decay on a fast time-scale relative to signal switching, retroactivity can be eliminated in the steady state for certain downstream systems $\mathcal{D}$ (as previously noted in~\cite{ghaemi} for a specific case). More generally, birth/death dynamics serves to buffer the concentration of $Z$ against the influence of $\mathcal{D}$, reducing retroactivity. 
However, such a buffering would incur substantial resource costs, requiring a high turnover of molecules or the establishment and maintenance of a very large buffer population.

We then consider the behaviour of certain catalytic circuits called push-pull motifs that can act as {insulators} $\mathcal{I}$, to explore whether they can reduce retroactivity at low cost. These insulators do not require a high production rate of signalling molecules, nor the establishment and maintenance of a large population of said molecules. Instead, the insulators consume fuel, typically by converting  ATP into ADP and inorganic phosphate.

We argue that coupling  $\mathcal{U}$ to $\mathcal{I}$ weakly  reduces both the retroactivity and fuel consumption. Moreover, in the steady-state signalling limit, it is often  possible to simultaneously reduce both retroactivity and fuel consumption to arbitrary  low levels, whilst  maintaining an approximately fixed signal propagation from $\mathcal{U}$ to the output of $\mathcal{D}$. Note that we do not claim that one can \textit{always} match an input-output curve with weaker coupling. In particular, motifs based on zero-order ultra-sensitivity~\cite{goldbeter1981amplified,berg2000fluctuations,ferrell2014ultrasensitivity,jithinraj2014zero} actually leverage retroactive effects. However, in such contexts retroactivity is a key ingredient of the system, rather than a nuisance to be eliminated. Additionally, in these cases  it would be incorrect to say that suppressing retroactivity requires more fuel consumption -- instead, suppressing retroactivity and fuel consumption simultaneously comes at the expense of signal alteration. 

Therefore it is in general possible to suppress retroactivity at low cost through insulation, and an engineer could design a signalling network with low energy consumption and low retroactivity. This observation still holds when the finite free energy associated with the breakdown of each ATP is explicitly modelled through microscopically reversible reactions. We note that unlike increasing the concentration of insulator molecules~\cite{ninfa,sontag_barton,shah2017signaling,legewie2008recurrent,dekel2005optimality}, which strongly influences the interactions of  both $\mathcal{U}$ and $\mathcal{D}$ with $\mathcal{I}$, there is no reason why changing the coupling of $\mathcal{I}$ to $\mathcal{U}$ at their mutual interface should make $\mathcal{I}$ more subject to ``retroactivity to the output" at its interface with downstream subsystems~\cite{ninfa}. This fact supports the approach of considering only the retroactivity on $\mathcal{U}$ in our analysis. However, the presence of unintended leak reactions limits the degree to which the coupling to the insulator can be weakened before signal transduction is compromised. 

In this work we have assumed that the signalling network reaches steady state. We have thus not considered its ability to respond to fast variation of the parameters of the upstream subsystem, as in some previous studies~\cite{ninfa,sontag_barton}. Tracking rapid variation in $\mathcal{U}$ is impossible if $\mathcal{U}$ only couples weakly to downstream subsystems since insulator molecules must undergo catalytic cycling on a time-scale comparable to the variation in $\mathcal{U}$ in order to propagate the time-varying signal. We therefore expect that, like robustness to leak reactions, the need to respond to time-varying signals will set a limit on how weak the coupling between $\mathcal{U}$ and $\mathcal{I}$ can be whilst retaining functional signalling. 

Note, however, that neither the constraints that arise from leak reactions nor those from signal-tracking imply that a high level of fuel consumption is necessary to suppress retroactivity. Rather, an alternative trade-off is suggested: retroactivity can be reduced at low free-energy cost, but at the expense of reduced response speed and robustness of the signalling pathway. Exploring this putative trade-off in more depth, and with more detailed models of chemical reactions, will be the subject of further work. 

From the perspective of understanding and engineering actual biochemical systems in an experimental context, relevant questions are: how weak can the coupling be in practice before signalling is disrupted, and is the principle of relatively weak coupling applicable in natural systems? In particular, if weaker coupling is used in biology to minimize fuel turnover, one would expect different circuits to find different optimal trade-offs. Circuits with many possible leak reactions, or which need to vary on a rapid time scale, will exhibit stronger coupling (faster fuel turnover) than others. Furthermore, our analysis may explain why an activation reaction that is known to be vital for cellular function nonetheless has a slow rate. 

When designing a synthetic signalling network, either from proteins or nucleic-acid based analogs, researchers could consider varying coupling strength to optimize performance, and indeed might consider different coupling strengths for different tasks. Importantly, making strong binding weaker by mutating a binding interface is relatively simple -- at least when compared to making an already-strong interface stronger. We note that it is important to make interactions with both the activating and deactivating catalysts weaker. 

From a fundamental biophysics perspective, our results emphasize an important and often mis-understood point. Catalytic circuits must be dissipative (consume fuel) in order to function. But given an inherently dissipative structure  it doesn't follow that an increased dissipation rate leads to better performance. Fundamentally, a fuel-consuming network structure is needed to ensure that catalysts overwhelmingly activate rather than deactivate their substrates (or vice versa), which is a question of relative reaction rates. The rate of dissipation, however, depends not only on these relative rates but also absolute rates, which may not help to improve circuit functionality \cite{baiesi2017}.

The above results are in stark contrast to the claims of Barton and Sontag in~\cite{sontag_barton} who analysed a similar system, but used different metrics to quantify retroactivity and considered a relatively rapid variation in the signal. Specifically, Barton and Sontag considered the same $\mathcal{D}$ and $\mathcal{I}$, but a $\mathcal{U}$ driven by a birth-death process with a sinusoidally varying birth rate $k(t)$.

They defined two metrics to quantify the amount of retroactivity in this system, namely the distortion and competition effect. The distortion captures the difference between the actual output $[C_{\text{real}}(t)]$ and the output $[C_{\text{ideal}}(t)]$ of a hypothetical system in which the downstream system responds to $Z$ as if binding were occurring, but the population of $Z$ is unaffected by these reactions (and thus there is no retroactivity). The distortion metric is given by
\begin{eqnarray}
\mathscr{D} = \frac{1}{\sigma_{[C_{\text{ideal}}]}}\left\langle\left|[C_{\text{ideal}}(t)] - [C_{\text{real}}(t)]\right|\right\rangle.
\end{eqnarray}
Here, $\sigma_{[C_{\text{ideal}}]}$ is the standard deviation of the ideal signal corresponding to the hypothetical system, and the angled brackets indicate an average over time. They also define the competition metric $\mathcal{C}$ as
\begin{eqnarray}
\mathscr{C} = \frac{1}{\sigma_{[C]}}\left\langle\left|\frac{\partial [C(t)]}{\partial [P'_{\text{tot}}]}\bigg|_{[P'_{\text{tot}}]=0}\right|\right\rangle,
\end{eqnarray}
where $[P^\prime_{\rm tot}]$ is the total concentration of a binding site for a second downstream subsystem $\mathcal{D}^\prime$. 

On the basis of these metrics, Barton and Sontag argued that \textit{producing better insulators requires substantial energy consumption}. However, these conclusions are a direct result of the particular choice of retroactivity metrics, which we believe are poorly-justified. Firstly, although the presence of retroactive terms in the dynamical equations does influence the output of a system, it is unclear why the deviation of the output from a particular hypothetical ``ideal" system should quantify retroactivity. For a start, one could write down other ``ideal" systems in which the retroactive terms were removed. But more importantly, the metric $\mathscr{D}$ doesn't quantify the back-action felt by  $\mathcal{U}$. For example, it is large if $\mathcal{U}$ is completely decoupled from any insulator and downstream network. Such a system does a poor job of propagating a signal, but doesn't exhibit retroactivity in any meaningful sense. 

The competition metric $\mathscr{C}$ comes closer to the spirit of retroactivity, in that it quantities the effect of one downstream subsystem   on another via $\mathcal{U}$. However, minimising $\mathscr{C}$ with respect to the parameters in $\mathcal{D}$ and/or $\mathcal{I}$, rather than the newly-added subsystem, does not minimise retroactivity due to the $\mathcal{I}+\mathcal{D}$ subsystem. Instead, it involves making $\mathcal{I}+\mathcal{D}$ insensitive to the presence of  the new downstream subsystem -- which can be achieved, for example, by coupling to $\mathcal{U}$ very strongly, so that the introduction of a new downstream system has essentially no effect. Such a design would be highly retroactive in the sense that $\mathcal{I}+\mathcal{D}$ strongly influences $\mathcal{U}$, but would have a low value of $\mathscr{C}$. 

We therefore believe that the ``optimal" systems found by Barton and Sontag do not minimise retroactivity. Instead they identified subsystems $\mathcal{I}$ that allow rapid tracking of $\mathcal{U}$, and are relatively insensitive to the introduction of parallel downstream subsystems, due to strong coupling between $\mathcal{U}$ and $\mathcal{I}$ that results in high fuel turnover.  We strongly advocate for the use of the retroactivity metric of  Del Vecchio {\it et al.} in future work \cite{ninfa}, to distinguish these distinct properties.

\section{Codes}
All the program codes for this manuscript can be accessed here: \url{http://www.imperial.ac.uk/principles-of-biomolecular-systems/contact--obtain-code-and-data/}
\section{Acknowledgements}
We thank Pieter Rein ten Wolde, Nick Jones and Manoj Gopalkrishnan for useful discussions. A.D. would like to thank the Math department at Imperial College London for funding through the ROTH scholarship. T.E.O. acknowledges funding through the Royal Society University Research Fellowship.
\bibliographystyle{unsrt}
\bibliography{Bibliography}

\begin{thebibliography}{10}

\bibitem{alon2007network}
U.~Alon.
\newblock Network motifs: theory and experimental approaches.
\newblock {\em Nat. Rev. Genet.}, 8(6):450--461, 2007.

\bibitem{Kashtan27092005}
N.~Kashtan and U.~Alon.
\newblock Spontaneous evolution of modularity and network motifs.
\newblock {\em Proc. Natl. Acad. Sci. U.S.A.}, 102(39):13773--13778, 2005.

\bibitem{hintze2008evolution}
A.~Hintze and C.~Adami.
\newblock Evolution of complex modular biological networks.
\newblock {\em PLoS. Comput. Biol.}, 4(2):e23, 2008.

\bibitem{Clune20122863}
J.~Clune, J.~Mouret, and H.~Lipson.
\newblock The evolutionary origins of modularity.
\newblock {\em Proc. R. Soc. Lond., B, Biol. Sci.}, 280(1755), 2013.

\bibitem{raff2000dissociability}
E.~Raff and R.~Raff.
\newblock Dissociability, modularity, evolvability.
\newblock {\em Evol. Dev}, 2(5):235--237, 2000.

\bibitem{Tran20130771}
T.~Tran and Y.~Kwon.
\newblock The relationship between modularity and robustness in signalling
  networks.
\newblock {\em J. R. Soc. Interface}, 10(88), 2013.

\bibitem{rorick2011protein}
M.~Rorick and G~Wagner.
\newblock Protein structural modularity and robustness are associated with
  evolvability.
\newblock {\em Genome Biol. Evol.}, 3:456--475, 2011.

\bibitem{kim2016robustness}
J.~Kim and K.~Cho.
\newblock Robustness analysis of network modularity.
\newblock {\em {IEEE} Trans. Control Netw. Syst}, 3(4):348--357, 2016.

\bibitem{hartwell:molecular}
L.~Hartwell, J.~Hopfield, S.~Leibler, and A.~Murray.
\newblock From molecular to modular cell biology.
\newblock {\em Nature}, 402:C47--C52, 1999.

\bibitem{Lauffenburger09052000}
D.~Lauffenburger.
\newblock Cell signaling pathways as control modules: Complexity for
  simplicity?
\newblock {\em Proc. Natl. Acad. Sci. U.S.A.}, 97(10):5031--5033, 2000.

\bibitem{purnick2009second}
P~Purnick and R~Weiss.
\newblock The second wave of synthetic biology: from modules to systems.
\newblock {\em Nat. Rev. Mol. Cell. Biol.}, 10(6):410--422, 2009.

\bibitem{andrianantoandro2006synthetic}
E.~Andrianantoandro, S.~Basu, D.~Karig, and R.~Weiss.
\newblock Synthetic biology: new engineering rules for an emerging discipline.
\newblock {\em Mol. Syst. Biol.}, 2(1), 2006.

\bibitem{SaezRodriguez2005619}
J.~Rodriguez, A.~Kremling, and E.~Gilles.
\newblock Dissecting the puzzle of life: modularization of signal transduction
  networks.
\newblock {\em Comput. Chem. Eng.}, 29(3):619 -- 629, 2005.

\bibitem{ventura2008hidden}
A.~Ventura, J.~Sepulchre, and S.~Merajver.
\newblock A hidden feedback in signaling cascades is revealed.
\newblock {\em PLoS. Comput. Biol.}, 4(3):e1000041, 2008.

\bibitem{ninfa}
D.~Del~Vecchio, A.~Ninfa, and E.~Sontag.
\newblock Modular cell biology: retroactivity and insulation.
\newblock {\em Mol. Syst. Biol.}, 4(1):161, 2008.

\bibitem{del2015biomolecular}
D.~Del~Vecchio and R.~Murray.
\newblock {\em Biomolecular feedback systems}.
\newblock Princeton University Press, 2015.

\bibitem{sontag_barton}
J.~Barton and E.~Sontag.
\newblock The {E}nergy {C}osts of {I}nsulators in {B}iochemical {N}etworks.
\newblock {\em Biophys. J.}, 104:1380--1390, 2013.

\bibitem{kim2010}
H.~Kim and H.~Sauro.
\newblock Fan-out in gene regulatory networks.
\newblock {\em J. Biol. Eng.}, 4(1):16, 2010.

\bibitem{shah2017signaling}
R.~Shah and D.~Del~Vecchio.
\newblock Signaling architectures that transmit unidirectional information
  despite retroactivity.
\newblock {\em bioRxiv}, page 111971, 2017.

\bibitem{barton_remark}
J.~Barton and E.~Sontag.
\newblock Remarks on the energy costs of insulators in enzymatic cascades.
\newblock {\em arXiv:1412.8065}.

\bibitem{ghaemi}
R.~Ghaemi and D.~Del~Vecchio.
\newblock Stochastic analysis of retroactivity in transcriptional networks
  through singular perturbation.
\newblock In {\em 2012 American Control Conference (ACC)}, pages 2731--2736,
  June 2012.

\bibitem{alon2006introduction}
U.~Alon.
\newblock {\em An Introduction to Systems Biology: Design Principles of
  Biological Circuits}.
\newblock CRC press, 2006.

\bibitem{Lemmon1996}
M.~A. Lemmon, K.~M. Ferguson, and J.~Schlessinger.
\newblock {PH domains: Diverse} sequences with a common fold recruit signaling
  molecules to the cell surface.
\newblock {\em Cell}, 85:621--624, 1996.

\bibitem{Qian2011}
L~Qian and E.~Winfree.
\newblock Scaling up digital circuit computation with {DNA} strand displacement
  cascades.
\newblock {\em Science}, 332:1196--1201, 2011.

\bibitem{Green2014}
A.~A. Green, P.~A. Silver, J.~Collins, and P.~Yin.
\newblock Toehold switches: De-novo-designed regulators of gene expression.
\newblock {\em Cell}, 159:925--939, 2014.

\bibitem{pantoja2015retroactivity}
L.~Pantoja-Hern{\'a}ndez and J.~Mart{\'\i}nez-Garc{\'\i}a.
\newblock Retroactivity in the context of modularly structured biomolecular
  systems.
\newblock {\em Front. Bioeng. Biotechnol.}, 3, 2015.

\bibitem{Mehta2016}
Pankaj Mehta, Alex~H. Lang, and David~J. Schwab.
\newblock Landauer in the age of synthetic biology: {Energy} consumption and
  information processing in biochemical networks.
\newblock {\em J. Stat. Phys.}, 162:1153--1166, 2016.

\bibitem{Ouldridge_PRX_2017}
T.~E. Ouldridge, C.~C. Govern, and P.~R. {ten Wolde}.
\newblock Thermodynamics of computational copying in biochemical systems.
\newblock {\em Phys. Rev. X.}, 7:021004, 2017.

\bibitem{robinson1997mitogen}
M.~Robinson and M.~Cobb.
\newblock Mitogen-activated protein kinase pathways.
\newblock {\em Curr. Opin. Cell. Biol.}, 9(2):180--186, 1997.

\bibitem{Samoilov15022005}
M.~Samoilov, S.~Plyasunov, and A.~Arkin.
\newblock Stochastic amplification and signaling in enzymatic futile cycles
  through noise-induced bistability with oscillations.
\newblock {\em Proc. Natl. Acad. Sci. U.S.A.}, 102(7):2310--2315, 2005.

\bibitem{ossareh2011long}
H.~Ossareh, A.~Ventura, S.~Merajver, and D.~Del~Vecchio.
\newblock Long {S}ignaling {C}ascades {T}end to {A}ttenuate {R}etroactivity.
\newblock {\em Biophys. J.}, 100(7):1617--1626, 2011.

\bibitem{mishra2014load}
Deepak Mishra, Phillip~M Rivera, Allen Lin, Domitilla Del~Vecchio, and Ron
  Weiss.
\newblock A load driver device for engineering modularity in biological
  networks.
\newblock {\em Nat. Biotechnol.}, 32(12):1268--1275, 2014.

\bibitem{sepulchre2012retroactive}
J.~Sepulchre, Sof{\'\i}a~D. Merajver, and A.~Ventura.
\newblock Retroactive signaling in short signaling pathways.
\newblock {\em PloS one}, 7(7):e40806, 2012.

\bibitem{catozzi2016signaling}
S.~Catozzi, J.~Di-Bella, A.~Ventura, and J.~Sepulchre.
\newblock Signaling cascades transmit information downstream and upstream but
  unlikely simultaneously.
\newblock {\em BMC Syst. Biol.}, 10(1):84, 2016.

\bibitem{wynn2011kinase}
M.~Wynn, A.~Ventura, J.~Sepulchre, H.~Garc{\'\i}a, and S.~Merajver.
\newblock Kinase inhibitors can produce off-target effects and activate linked
  pathways by retroactivity.
\newblock {\em BMC Syst. Biol.}, 5(1):156, 2011.

\bibitem{feliu2012algebraic}
E.~Feliu, M.~Knudsen, L.~Andersen, and C.~Wiuf.
\newblock An algebraic approach to signaling cascades with n layers.
\newblock {\em Bull. Math. Biol.}, 74(1):45--72, 2012.

\bibitem{Govern09122014}
C.~Govern and P.~ten Wolde.
\newblock Optimal resource allocation in cellular sensing systems.
\newblock {\em Proc. Natl. Acad. Sci. U.S.A.}, 111(49):17486--17491, 2014.

\bibitem{huang1996ultrasensitivity}
C.~Huang and J.~Ferrell.
\newblock Ultrasensitivity in the mitogen-activated protein kinase cascade.
\newblock {\em Proc. Natl. Acad. Sci. U.S.A.}, 93(19):10078--10083, 1996.

\bibitem{Sontag2011}
E.~Sontag.
\newblock {\em Modularity, Retroactivity, and Structural Identification}, pages
  183--200.
\newblock 2011.

\bibitem{jayanthi2013retroactivity}
S.~Jayanthi, K.~Nilgiriwala, and D.~Del~Vecchio.
\newblock Retroactivity controls the temporal dynamics of gene transcription.
\newblock {\em ACS Synth. Biol.}, 2(8):431--441, 2013.

\bibitem{stock2000two}
A.~Stock, V.~Robinson, and P.~Goudreau.
\newblock Two-component signal transduction.
\newblock {\em Annu. Rev. Biochem.}, 69(1):183--215, 2000.

\bibitem{novick1950description}
A.~Novick and L.~Szilard.
\newblock Description of the chemostat.
\newblock {\em Science}, 112(2920):715--716, 1950.

\bibitem{aris1965prolegomena}
R.~Aris.
\newblock Prolegomena to the rational analysis of systems of chemical
  reactions.
\newblock {\em Arch. Rational Mech. Anal.}, 19(2):81--99, 1965.

\bibitem{aris1968prolegomena}
R.~Aris.
\newblock Prolegomena to the rational analysis of systems of chemical reactions
  ii. some addenda.
\newblock {\em Arch. Rational Mech. Anal.}, 27(5):356--364, 1968.

\bibitem{feinberg1989necessary}
M.~Feinberg.
\newblock Necessary and sufficient conditions for detailed balancing in mass
  action systems of arbitrary complexity.
\newblock {\em Chem. Eng. Sci.}, 44(9):1819--1827, 1989.

\bibitem{gunawardena2003chemical}
J.~Gunawardena.
\newblock Chemical reaction network theory for in-silico biologists.
\newblock {\em Notes available for download at http://vcp. med. harvard.
  edu/papers/crnt. pdf}, 2003.

\bibitem{horn1972general}
F.~Horn and R.~Jackson.
\newblock General mass action kinetics.
\newblock {\em Arch. Rational Mech. Anal.}, 47(2):81--116, 1972.

\bibitem{bender2014introduction}
D.~Bender.
\newblock {\em Introduction to nutrition and metabolism}.
\newblock CRC Press, 2014.

\bibitem{stadtman1977superiority}
E.~Stadtman and P.~Chock.
\newblock Superiority of interconvertible enzyme cascades in metabolic
  regulation: analysis of monocyclic systems.
\newblock {\em Proc. Natl. Acad. Sci. U.S.A.}, 74(7):2761--2765, 1977.

\bibitem{goldbeter1981amplified}
A.~Goldbeter and D.~Koshland.
\newblock An amplified sensitivity arising from covalent modification in
  biological systems.
\newblock {\em Proc. Natl. Acad. Sci. U.S.A.}, 78(11):6840--6844, 1981.

\bibitem{legewie2008recurrent}
S.~Legewie, H.~Herzel, H.~Westerhoff, and N.~Bl{\"u}thgen.
\newblock Recurrent design patterns in the feedback regulation of the mammalian
  signalling network.
\newblock {\em Mol. Syst. Biol.}, 4(1):190, 2008.

\bibitem{dekel2005optimality}
E.~Dekel and U.~Alon.
\newblock Optimality and evolutionary tuning of the expression level of a
  protein.
\newblock {\em Nature}, 436(7050):588, 2005.

\bibitem{tenWolde2016}
P.~R. ten Wolde, N.~B. Becker, T.~E. Ouldridge, and A.~Mugler.
\newblock Fundamental limits to cellular sensing.
\newblock {\em J. Stat. Phys.}, 162:1395--1424, 2016.

\bibitem{Lewis01031925}
G.~Lewis.
\newblock A new principle of equilibrium.
\newblock {\em Proc. Natl. Acad. Sci. U.S.A.}, 11(3):179--183, 1925.

\bibitem{Ouldridge_review_2017}
T.~E. Ouldridge.
\newblock The importance of thermodynamics for molecular systems, and the
  importance of molecular systems for thermodynamics.
\newblock {\em arXiv:1702.00360}.

\bibitem{zubay1998biochemistry}
G.~Zubay.
\newblock {\em Biochemistry}.
\newblock Brown Publishers, US, 1998.

\bibitem{friedlander2016intrinsic}
T.~Friedlander, R.~Prizak, C.~Guet, N.~Barton, and G.~Tka{\v{c}}ik.
\newblock Intrinsic limits to gene regulation by global crosstalk.
\newblock {\em Nat. Commun.}, 7, 2016.

\bibitem{Swain2002}
P.~S. Swain and E.~D. Siggia.
\newblock The role of proofreading in signal transduction specificity.
\newblock {\em Biophys. J.}, 82:2928--2933, 2002.

\bibitem{Ouldridge_proofreading_2014}
T.~E. Ouldridge and P.~R. ten Wolde.
\newblock The robustness of proofreading to crowding-induced
  pseudo-processivity in the {MAPK} pathway.
\newblock {\em Biophys. J.}, 107:2425 -- 2435, 2014.

\bibitem{berg2000fluctuations}
O.~Berg, J.~Paulsson, and M.~Ehrenberg.
\newblock Fluctuations and quality of control in biological cells: zero-order
  ultrasensitivity reinvestigated.
\newblock {\em Biophys. J.}, 79(3):1228--1236, 2000.

\bibitem{ferrell2014ultrasensitivity}
J.~Ferrell and S.~Ha.
\newblock Ultrasensitivity part {I}: Michaelian responses and zero-order
  ultrasensitivity.
\newblock {\em Trends Biochem. Sci.}, 39(10):496--503, 2014.

\bibitem{jithinraj2014zero}
P.~Jithinraj, U.~Roy, and M.~Gopalakrishnan.
\newblock Zero-order ultrasensitivity: A study of criticality and fluctuations
  under the total quasi-steady state approximation in the linear noise regime.
\newblock {\em J. Theor. Biol.}, 344:1--11, 2014.

\bibitem{baiesi2017}
M.~Baiesi and C.~Maes.
\newblock Life efficiency does not always increase with the dissipation rate.
\newblock {\em arXiv:1707.09614}, 2017.

\end{thebibliography}
\appendix

\section{Appendix}
In what follows, we will assume that we start with a fixed amount of transcription factor $[Z_{\text{tot}}]$, promoter $[P_{\text{tot}}]$, kinase $[Z_{\text{tot}}]$ and phosphatase $[Y_{\text{tot}}]$ unless specified otherwise.
\subsection{Analytics for different Z dynamics}\label{sec:analytics_Z_dynamics}
In this section we give analytical results for the steady-state dynamics corresponding to three different $\mathcal{U}$ subsystems discussed in Section~\ref{sec:direct_binding} in the main text.
\begin{enumerate}
\item\textbf{Fixed amount of $Z$:} 
\begin{align}
Z + P\xrightleftharpoons[k_{\text{off}}]{k_{\text{on}}}C\nonumber
\end{align}
In this case, we have the conservation laws $[Z^{\rm ss}] + [C^{\rm ss}] = [Z_{\text{tot}}]$ and $[P^{\rm ss}] + [C^{\rm ss}] = [P_{\text{tot}}]$. Solving for steady-state, we get $k_{\text{on}}[Z^{\rm ss}]([P_{\text{tot}}] - [Z_{\text{tot}}] + [Z^{\rm ss}])= k_{\text{off}}([Z_{\text{tot}}] - [Z^{\rm ss}])$ implying that 
\begin{align*}
[Z^{\rm ss}] = \frac{-\lambda\pm\sqrt{\lambda^2 + 4k_{\text{off}}k_{\text{on}}[Z_{\text{tot}}]}}{2k_{\text{on}}},
\end{align*}
where $\lambda = k_{\text{off}} + k_{\text{on}}([P_{\text{tot}}] - [Z_{\text{tot}}])$. We choose the solution that makes physical sense for a given set of parameters i.e. the solution that satisfies $[Z^{\rm ss}]\geq 0$ and $[Z^{\rm ss}]\leq [Z_{\text{tot}}]$. The metric for retroactivity translates to 
\begin{eqnarray*}
\mathcal{R}= \left| 1 - \frac{[Z^{\rm ss}]}{[Z_{\mathcal{D,I}  \rightarrow \emptyset}^{\rm ss}]}\right| = 1 - \frac{[Z^{\rm ss}]}{[Z_{\text{tot}}]}
\end{eqnarray*}
  
\item\textbf{Constant birth/death dynamics:} 
\begin{align}
\phi &\xrightleftharpoons[\delta]{k}Z^*\nonumber\\
Z + P &\xrightleftharpoons[k_{\text{off}}]{k_{\text{on}}}C\nonumber
\end{align}
In this case, solving for steady-state we get $k - \delta[Z^{\rm ss}] - k_{\text{on}}[Z^{\rm ss}]([P_{\text{tot}}] - [C^{\rm ss}]) + k_{\text{off}}[C^{\rm ss}]=0$ and $k_{\text{on}}[Z^{\rm ss}]([P_{\text{tot}}] - [C^{\rm ss}]) - k_{\text{off}}[C^{\rm ss}]=0$. Therefore $[Z^{\rm ss}] =\frac{k}{\delta}$. In addition, note that $[Z_{\mathcal{D,I}  \rightarrow \emptyset}^{\rm ss}]=\frac{k}{\delta}$ implying that the retroactivity metric is
\begin{eqnarray}
\mathcal{R}= \left| 1 - \frac{[Z^{\rm ss}]}{[Z_{\mathcal{D,I}  \rightarrow \emptyset}^{\rm ss}]}\right| = 0.
\end{eqnarray}
\item\textbf{Active/Inactive forms of $Z$:} 
\begin{align}
Z_0 &\xrightleftharpoons[k_{\text{in}}]{k_{\text{ac}}}Z\nonumber\\
Z + P &\xrightleftharpoons[k_{\text{off}}]{k_{\text{on}}}C\nonumber
\end{align}
In this case, we start with a fixed amount of total $Z$, say $[Z'_{\text{tot}}]$. Solving for steady-state, we get $k_{\text{ac}}([Z'_{\text{tot}}] - [Z^{\rm ss}] - [C^{\rm ss}]) = k_{\text{in}}[Z^{\rm ss}]$ and $k_{\text{on}}[Z^{\rm ss}]([p_{\text{tot}}] - [C^{\rm ss}]) = k_{\text{off}}[C^{\rm ss}]$ implying that 
\begin{align*}
\resizebox{0.45\textwidth}{!}{$[Z^{\rm ss}] = \frac{-\mu\pm\sqrt{\mu^2 + 4k_{\text{active}}(k_{\text{ac}} + k_{\text{in}}) k_{\text{off}}k_{\text{on}}[Z'_{\text{tot}}]}}{2(k_{\text{ac}}+ k_{\text{in}})k_{\text{on}}}$}
\end{align*}
where
\begin{eqnarray*}
\resizebox{0.45\textwidth}{!}{$\mu = k_{\text{in}}k_{\text{off}} + k_{\text{ac}}(k_{\text{off}} + k_{\text{on}}([P_{\text{tot}}] - [Z'_{\text{tot}}]))$}
\end{eqnarray*}
As in case $1$, we choose only those solutions that make physical sense i.e. those which satisfy $[Z^{\rm ss}]\geq 0$ and $[Z^{\rm ss}]\leq [Z'_{\text{tot}}]$. Note that $k_{\text{in}}[Z_{\mathcal{D,I}  \rightarrow \emptyset}^{\rm ss}]= k_{\text{ac}}([Z'_{\text{tot}}] - [Z_{\mathcal{D,I}  \rightarrow \emptyset}^{\rm ss}])$. Therefore,
\begin{eqnarray}
[Z_{\mathcal{D,I}\rightarrow \emptyset}^{\rm ss}] = \frac{k_{\text{ac}}}{k_{\text{ac}} + k_{\text{in}}}[Z'_{\text{tot}}]
% = [Z_{\text{tot}}]
\end{eqnarray}
Choosing $[Z_{\mathcal{D,I}  \rightarrow \emptyset}^{\rm ss}] = [Z_{\rm tot}]$ allows the system to be compared sensibly to other designs of $\mathcal{U}$ with the same behaviour in this limit. In this case, the metric for retroactivity translates to  
\begin{eqnarray}
\mathcal{R}= \left| 1 - \frac{[Z^{\rm ss}]}{[Z_{\mathcal{D,I}  \rightarrow \emptyset}^{\rm ss}]}\right| = 1 - \frac{[Z^{\rm ss}]}{[Z_{\text{tot}}]}.
\end{eqnarray}
\end{enumerate}
\subsection{Effect of decreasing the coupling to the push-pull}\label{low_retroactivity}
Recall the microscopically reversible push-pull motif from the main text:
\begin{align}
Z+X &\xrightleftharpoons[\beta_2]{\beta_1}C_1\xrightleftharpoons[\frac{\epsilon k_1\beta_1}{\beta_2}]{k_1} X^* + Z \nonumber\\
Y+X^*&\xrightleftharpoons[\alpha_2]{\alpha_1}C_2\xrightleftharpoons[\frac{\epsilon k_2\alpha_1}{\alpha_2}]{k_2} X + Y \nonumber\\
X^*+p&\xrightleftharpoons[k_{\text{off}}]{k_{\text{on}}}C.
\end{align}
Here, $0\leq\epsilon\leq 1$ is the parameter that defines the degree of microscopic reversibility. We analyse the effect of repeatedly reducing the coupling to the push-pull motif to match a given input/output curve. We show that the steady-state output of a push-pull motif is a function of the ratio $\frac{\beta_1}{\alpha_1}$ and not $\alpha_1$ and $\beta_1$ individually, in the low retroactivity limit. Further the retroactivity and power is directly proportional to $\beta_1$ at fixed ratio $\frac{\beta_1}{\alpha_1}$. Our analysis is divided into two cases:
\begin{enumerate}
\item\textbf{Microscopically irreversible limit:} A push-pull motif coupled to fuel with an infinite free energy corresponds to case $\epsilon=0$. Specifically, we have the following network:
\begin{align}
Z+X &\xrightleftharpoons[\beta_2]{\beta_1}C_1\xrightarrow{k_1} X^* + Z \nonumber\\
Y+X^*&\xrightleftharpoons[\alpha_2]{\alpha_1}C_2\xrightarrow{k_2} X + Y \nonumber\\
X^*+p&\xrightleftharpoons[k_{\text{off}}]{k_{\text{on}}}
\end{align}
As we reduce the coupling to the push-pull motif by making $\alpha_1$ and $\beta_1$ sufficiently small, one can approximately ignore sequestration into complexes relative to $[C^{\rm ss}]$, $[X^{\rm ss}]$ and $[X^{*{\rm ss}}]$ and the network essentially boils down to the following:
\begin{align}
X&\xrightleftharpoons[\frac{\alpha_1k_2[Y_{\text{tot}}]}{k_2 + \alpha_2}]{\frac{k_1\beta_1[Z_{\text{tot}}]}{\beta_2+k_1}}  X^* \nonumber\\
X^*+P&\xrightleftharpoons[k_{\text{off}}]{k_{\text{on}}}C.
\end{align}
Solving for steady-state, we get
\begin{eqnarray}
\resizebox{0.35\textwidth}{!}{[$X^{\rm ss}]\frac{k_1\beta_1[Z_{\text{tot}}]}{\beta_2 + k_1}=[X^{\rm *ss}]\frac{\alpha_1k_2[Y_{\text{tot}}]}{k_2 + \alpha_2}$}
\end{eqnarray}
and 
\begin{eqnarray}
\resizebox{0.35\textwidth}{!}{$k_{\text{on}}[X^{\rm *ss}]([P_{\text{tot}}]-[C^{\rm ss}])=k_{\text{off}}[C^{\rm ss}]$}
\end{eqnarray}
with the conservation relation $[X^{\rm ss}] + [X^{\rm *ss}] + [C^{\rm ss}]=[X_{\text{tot}}]$. Therefore, we have 
\begin{align}\label{eq:no_sequestration_irreversible}
\resizebox{0.35\textwidth}{!}{$[C^{\rm ss}] = \frac{f(r)\pm\sqrt{f^2(r) - 4r^2{k^2_{\text{on}}}[X_{\text{tot}}][P_{\text{tot}}]}}{2rk_{\text{on}}}$}
\end{align}
where $r = \frac{\frac{k_1\beta_1[Z_{\text{tot}}]}{\beta_2+k_1}}{\frac{\alpha_1k_2[Y_{\text{tot}}]}{k_2 + \alpha_2}}$ and $f(r)=k_{\text{off}} + r(k_{\text{on}}([P_{\text{tot}}] + [X_{\text{tot}}]) + k_{\text{off}})$, justifying Equation~\ref{eq:irreversible_output} in the main text. It follows that both $[X^{\rm ss}]$ and $[X^{*\rm ss}]$ are functions of $\alpha_1$ and $\beta_1$ through the ratio $r=\frac{\beta_1}{\alpha_1}$.\\
Solving for steady-state of $[C_1]$, we get $\beta_1[Z^{\rm ss}][X^{*\rm ss}] - (\beta_2 + k_1)[C_1^{\rm ss}] = 0$, implying that $[C_1^{\rm ss}] = \frac{\beta_1 [Z^{\rm ss}][X^{*\rm ss}]}{(\beta_2 + k_1)}$. As a consequence $C_1^{\rm ss}\propto[Z^{\rm ss}]\beta_1$ at fixed $r$. This implies that $\mathcal{R}=\frac{[C_1^{\rm ss}]}{[Z_{\text{tot}}]}\propto\beta_1({1 + \beta_1})^{-1}\approx\beta_1$ for sufficiently small $\beta_1$ and flux $\Psi=k_1[C_1^{\rm ss}]\propto\beta_1$ at fixed $r$,  justifying Equation~\ref{eq:irreversible_retroactivity_flux} in the main text.
\item\textbf{Finite free energy of fuel molecules:} Explicitly incorporating the presence of microscopically reversible reactions in the push-pull motif corresponds to the case $0<\epsilon\leq 1$. As in the case of infinite free energy, making $\alpha_1$ and $\beta_1$ sufficiently small amounts to neglecting sequestration into complexes relative to $[C^{\rm ss}]$, $[X^{\rm ss}]$ and $[X^{*{\rm ss}}]$, giving the following network:
\begin{align}
X &\xrightleftharpoons[\frac{\alpha_1k_2[Y_{\text{tot}}]}{k_2 + \alpha_2} + \epsilon\frac{k_1\beta_1[Z_{\text{tot}}]}{\beta_2+k_1}]{\frac{k_1\beta_1[Z_{\text{tot}}]}{\beta_2+k_1} + \epsilon\frac{\alpha_1k_2[Y_{\text{tot}}]}{k_2 + \alpha_2}}  X^* \nonumber\\
X^*+p &\xrightleftharpoons[k_{\text{off}}]{k_{\text{on}}}C\nonumber
\end{align}
Solving for steady-state, we get
\begin{eqnarray*}
\resizebox{0.44\textwidth}{!}{$X^{\rm ss}\bigg(\frac{k_1\beta_1[Z_{\text{tot}}]}{\beta_2+k_1} + \epsilon\frac{\alpha_1k_2[Y_{\text{tot}}]}{k_2 + \alpha_2}\bigg)=[X^{\rm *ss}]\bigg(\frac{\alpha_1k_2[Y_{\text{tot}}]}{k_2 + \alpha_2} + \epsilon\frac{k_1\beta_1[Z_{\text{tot}}]}{\beta_2+k_1}\bigg)$}
\end{eqnarray*}
and
\begin{eqnarray*}
k_{\text{on}}[X^{\rm *ss}]([P_{\text{tot}}]-[C^{\rm ss}])=k_{\text{off}}[C^{\rm ss}]
\end{eqnarray*}
with the conservation relation $[X^{\rm ss}] + [X^{\rm *ss}] + [C^{\rm ss}] = [X_{\text{tot}}]$. Let $r = \frac{\frac{k_1\beta_1[Z_{\text{tot}}]}{\beta_2+k_1}}{\frac{\alpha_1k_2[Y_{\text{tot}}]}{k_2 + \alpha_2}}$ and $r' = \frac{r + \epsilon}{r\epsilon + 1}$. Therefore, we have 
\begin{align}\label{eq:no_sequestration_reversible}
\resizebox{0.4\textwidth}{!}{[$C^{\rm ss}] = \frac{f(r')\pm\sqrt{f^2(r') - 4r'^2{k^2_{\text{on}}}[X_{\text{tot}}][P_{\text{tot}}]}}{2r'k_{\text{on}}}$}
\end{align}
where $f(r')=k_{\text{off}} + r'(k_{\text{on}}([P_{\text{tot}}] + [X_{\text{tot}}]) + k_{\text{off}})$. 
\end{enumerate}
Solving for $[C^{\rm ss}_1]$, we get $\beta_1[Z^{\rm ss}][X^{*\rm ss}] - (\beta_2 + k_1)[C_1^{\rm ss}] + \frac{\epsilon k_1\beta_1}{\beta_2}[X^{\rm *ss}]= 0$, implying that 
\begin{align}
[C_1^{\rm ss}] = \frac{\beta_1[Z^{\rm ss}][X^{*\rm ss}] + \frac{\epsilon k_1\beta_1}{\beta_2}[X^{\rm *ss}]}{(\beta_2 + k_1)}.
\end{align}
Since both $[X^{\rm ss}]$ and $[X^{\rm *ss}]$ depend only on the ratio $r'$, we get that $\mathcal{R}=\frac{[C_1^{\rm ss}]}{[Z_{\text{tot}}]}\propto\beta_1({1 + \beta_1})^{-1}\approx\beta_1$ for sufficiently small $\beta_1$ at fixed $r'$ and power $w=\Psi\Delta G_{\rm ATP}\propto\beta_1$ justifying equation~\ref{eq:reversible_retractivity_power} in the main text.

\subsection{Effectively irreversible push-pull motifs}\label{sec:4kT_criterion}
\begin{figure*}[t]
  \subfigure{\includegraphics[scale=0.15]{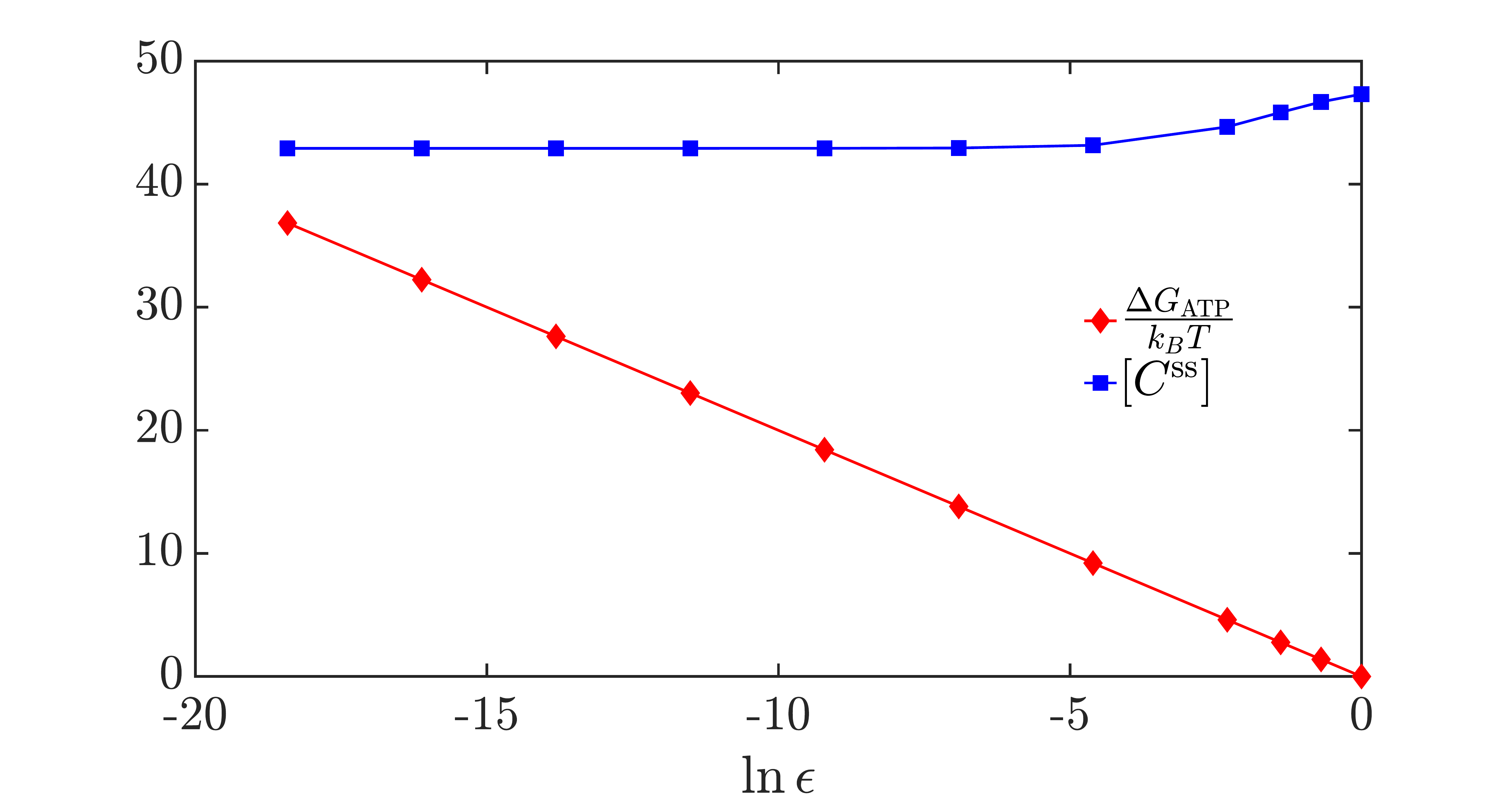}}\hspace{6mm}
  \subfigure{\includegraphics[scale=0.15]{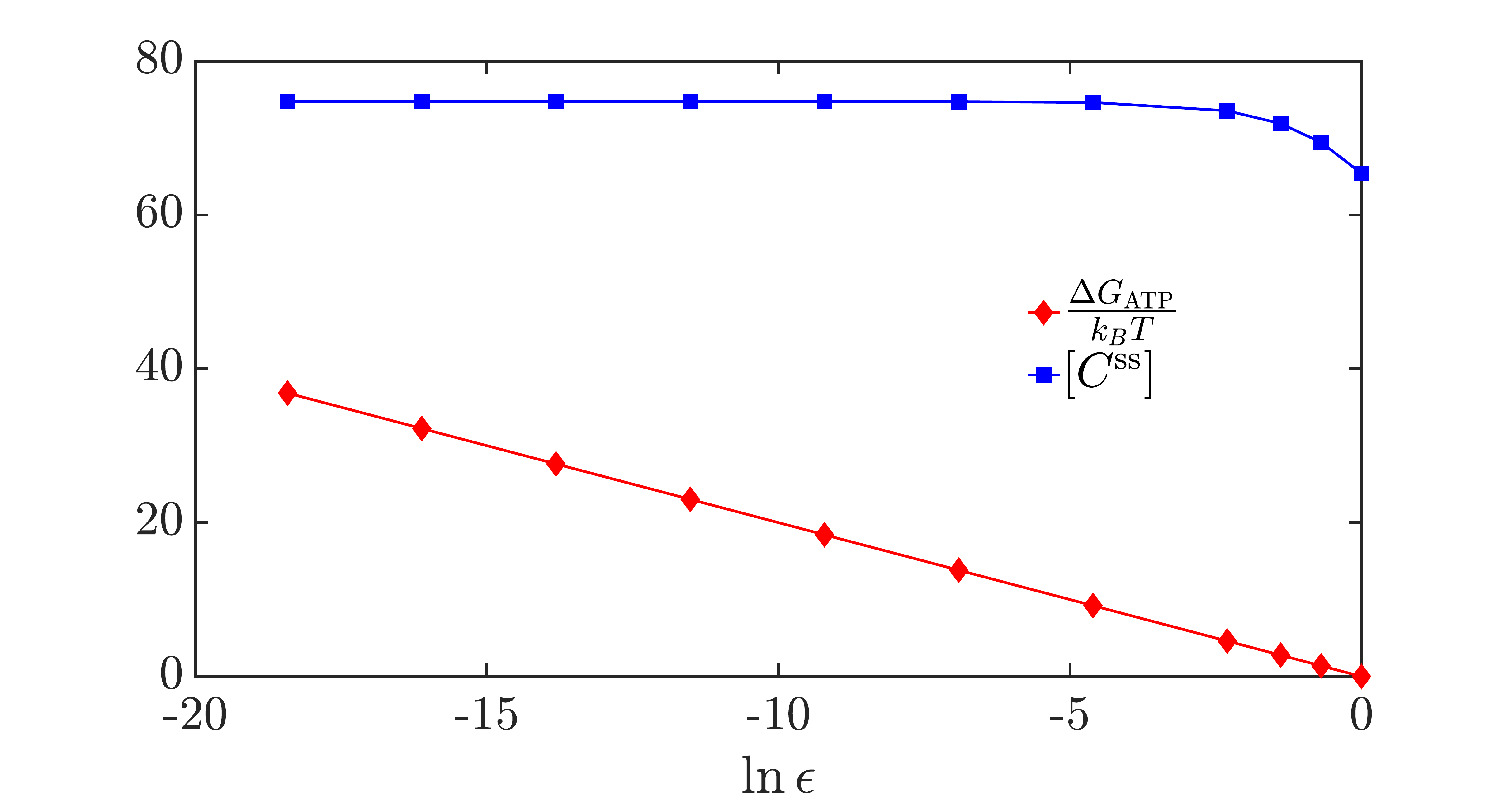}}
  \caption{For a push-pull network consuming free energy in excess of $ -2k_BT \ln \epsilon \sim4k_BT$, the presence of microscopically reversible reactions has negligible effect on the steady-state output as demonstrated for two specific systems. Parameters used for the model: $ a)\hspace{1mm}[X_{\text{tot}}]=[Y_{\text{tot}}]=[Z_{\text{tot}}] = 50,[P_{\text{tot}}] = 100,\alpha_1=\beta_1 = 0.1,\alpha_2 = \beta_2 = k_2 = k_{\text{on}} = k_{\text{off}} = 10,k_1=1$.\hspace{1mm} $b)\hspace{1mm}[X_{\text{tot}}]=200,[Y_{\text{tot}}]=[Z_{\text{tot}}]=100,[P_{\text{tot}}] = 100,\alpha_1=\beta_1=\alpha_2 = \beta_2 = k_1=k_2=1, k_{\text{on}} = k_{\text{off}} = 10$.}\label{fig:reversible_irreversible}
\end{figure*}

We show that push-pull networks consuming free energy per cycle beyond a certain threshold are essentially equivalent to those without the microscopically reversible reactions, for the purposes of the steady-state concentrations. In our system, this threshold is $\sim4k_BT$. Figure~\ref{fig:reversible_irreversible} illustrates this fact for certain sets of parameters. Recall from the main text that the free energy of a push-pull motif having microscopically reversible reactions is given by ${\Delta G}_{\rm ATP}=2k_BT\ln\epsilon$.

\subsection{Retroactivity and rate of free-energy consumption for randomly parameterised push-pull motifs }\label{subsec:example_finite_free_energy}
In this section we show that the results presented in Fig.~\ref{fig:retroactivity_energy_reversibility} of the main text -- namely that it is possible to reproduce an input-output relation at weaker coupling, thereby reducing retroactivity and free energy consumption, is true for a good proportion of randomly generated systems.
{Push-pull motifs having infinite free energy are a limiting case of generic push-pull networks possessing microscopically reversible reactions. In fact, as noted earlier, having an infinite free energy corresponds to putting $\epsilon=0$, where $\epsilon$ is the parameter that quantifies the amount of microscopic reversibility. It therefore suffices to consider push-pull motifs having finite free energy. We repeat the plots of Fig.~\ref{fig:retroactivity_energy_reversibility} from the main text, using randomly generated parameters. Specifically, we consider 10 systems randomly chosen from parameter distributions 
\begin{itemize}
\item $X_{\rm tot},Y_{\rm tot}\sim{\rm U}[1,200]$ \\
\item $P_{\rm tot}\sim{\rm U}[1,X_{\rm tot}]$ \\
\item $\alpha_1,\beta_1,k_{\rm on}\sim 10^{{\rm U}[-2,0]}$ \\
\item $\alpha_2,\beta_2,k_1,k_2,k_{\rm off}\sim 10^{{\rm U}[-1,1]}$ \\
\item $\epsilon\sim 10^{{\rm U}[-3,-1]}$,
\end{itemize}
Here, $U$ indicates a uniform distribution and all samples are independent. The results are plotted in the following, demonstrating that often (although not always) it is possible to get a very close match to the input-output curve at weaker coupling. Moreover, even when the matching of the input-output curve is only moderate, both retroactivity and fuel consumption still decrease.  
}
\begin{figure*}
\centering
\includegraphics[scale=0.46]{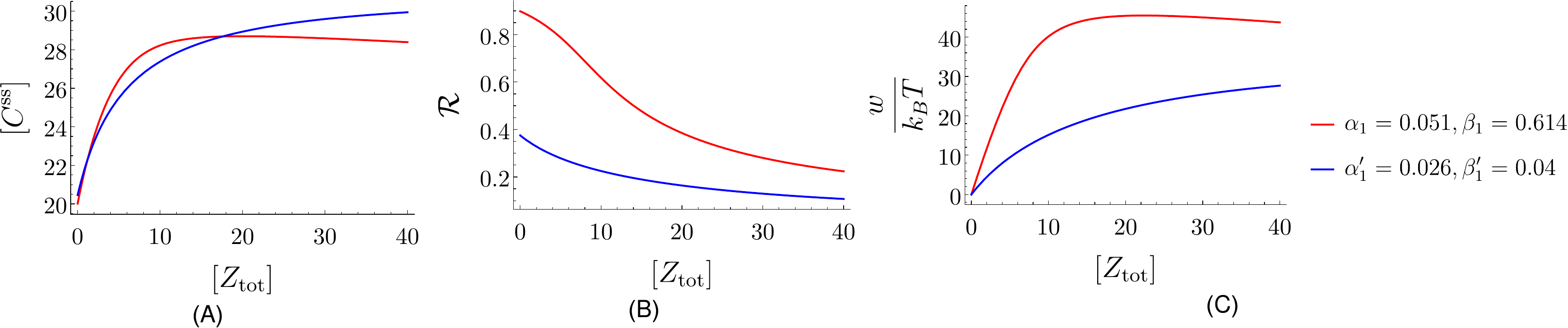}
\caption{$[X_{\text{tot}}]=41.198,[Y_{\text{tot}}]=108.948,[P_{\text{tot}}]=36.176,\alpha_1=0.051,\alpha_2=0.897,\beta_1=0.614,\beta_2=0.291,k_1=0.950,k_2=3.184,k_{\text{on}}=0.517,k_{\text{off}}=0.203,\epsilon=0.026$.}
\label{fig:reversible_1}
\end{figure*}
\begin{figure*}
\centering
\includegraphics[scale=0.46]{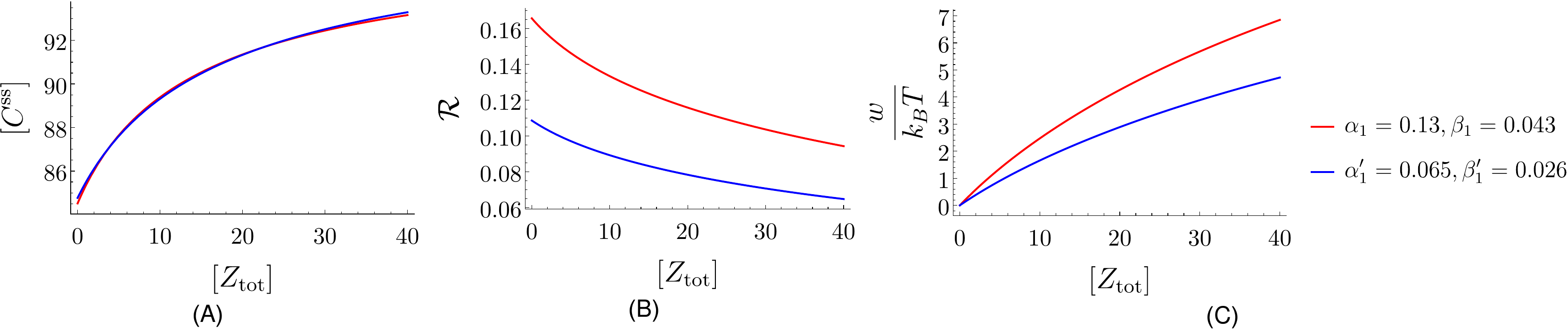}
\caption{$[X_{\text{tot}}]=124.747,[Y_{\text{tot}}]=5.810,[P_{\text{tot}}]=101.106,\alpha_1=0.126,\alpha_2=0.177,\beta_1=0.043,\beta_2=6.802,k_1=0.391,k_2=1.600,k_{\text{on}}=0.366,k_{\text{off}}=0.215,\epsilon=0.065$.}
\label{fig:reversible_2}
\end{figure*}
\begin{figure*}
\centering
\includegraphics[scale=0.46]{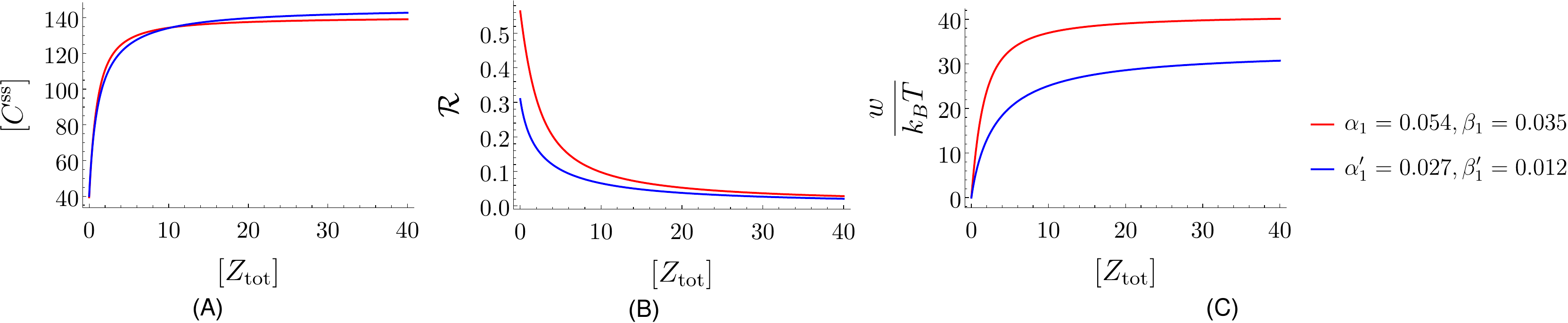}
\caption{$[X_{\text{tot}}]=182.861,[Y_{\text{tot}}]=57.132,[P_{\text{tot}}]=171.82,\alpha_1=0.054,\alpha_2=0.137,\beta_1=0.035,\beta_2=0.283,k_1=3.383,k_2=0.105,k_{\text{on}}=0.432,k_{\text{off}}=0.646,\epsilon=0.027$.}
\label{fig:reversible_3}
\end{figure*}
\begin{figure*}
\centering
\includegraphics[scale=0.46]{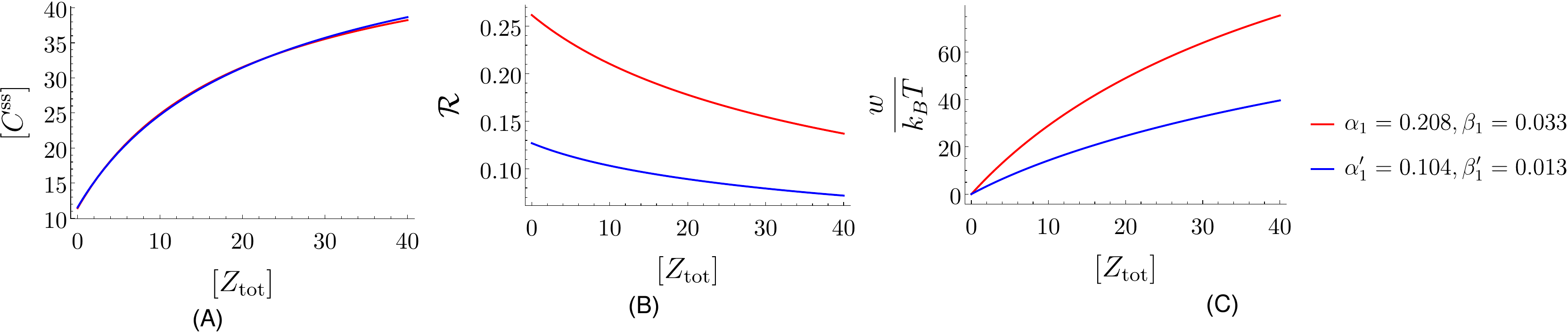}
\caption{$[X_{\text{tot}}]=95.631,[Y_{\text{tot}}]=147.689,[P_{\text{tot}}]=72.11,\alpha_1=0.208,\alpha_2=9.877,\beta_1=0.033,\beta_2=6.076,k_1=1.450,k_2=0.766,k_{\text{on}}=0.418,k_{\text{off}}=1.559,\epsilon=0.104$.}
\label{fig:reversible_4}
\end{figure*}
\begin{figure*}
\centering
\includegraphics[scale=.46]{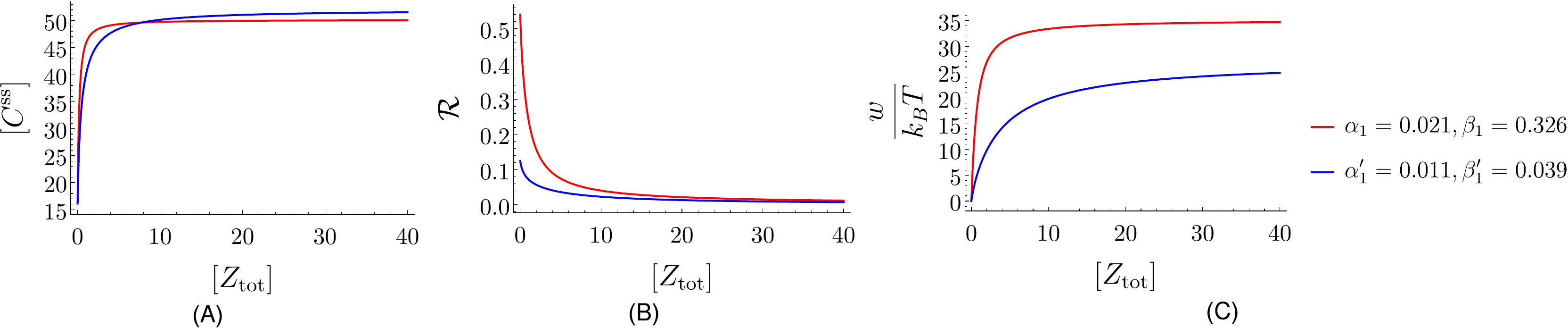}
\caption{$[X_{\text{tot}}]=84.151,[Y_{\text{tot}}]=100.826,[P_{\text{tot}}]=56.871,\alpha_1=0.021,\alpha_2=0.438,\beta_1=0.326,\beta_2=9.358,k_1=8.781,k_2=0.145,k_{\text{on}}=0.323,k_{\text{off}}=0.389,\epsilon=0.011$.}
\label{fig:reversible_5}
\end{figure*}
\begin{figure*}
\centering
\includegraphics[scale=.46]{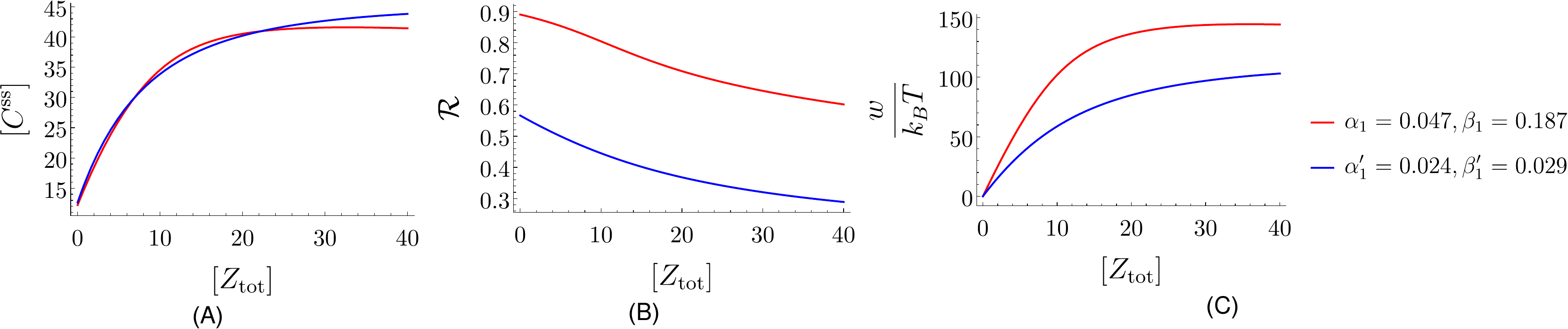}
\caption{$[X_{\text{tot}}]=138.883,[Y_{\text{tot}}]=67.044,[P_{\text{tot}}]=67.68,\alpha_1=0.047,\alpha_2=0.942,\beta_1=0.187,\beta_2=0.264,k_1=2.308,k_2=0.717,k_{\text{on}}=0.041,k_{\text{off}}=0.822,\epsilon=0.024$.}
\label{fig:reversible_6}
\end{figure*}
\begin{figure*}
\centering
\includegraphics[scale=.46]{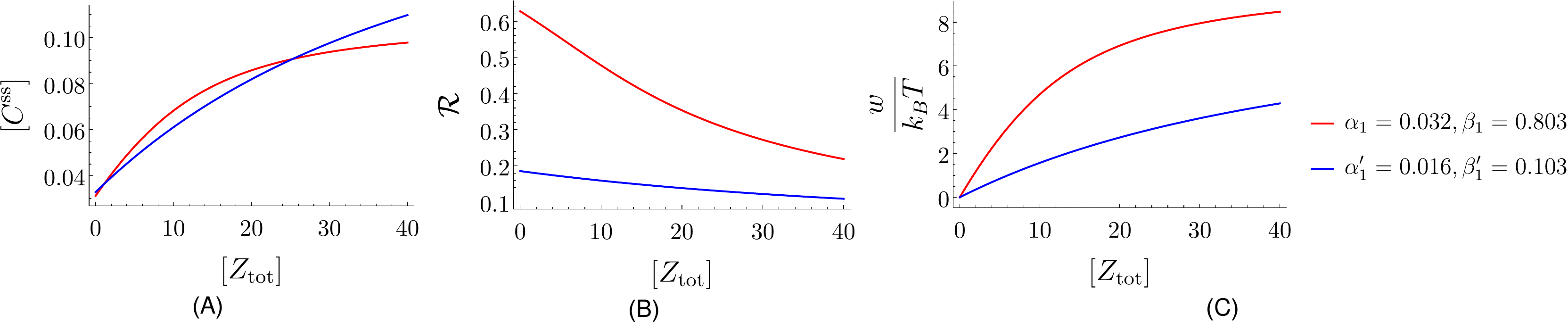}
\caption{$[X_{\text{tot}}]=14.304,[Y_{\text{tot}}]=58.104,[P_{\text{tot}}]=6.824,\alpha_1=0.032,\alpha_2=0.503,\beta_1=0.803,\beta_2=5.777,k_1=0.145,k_2=0.671,k_{\text{on}}=0.069,k_{\text{off}}=6.124,\epsilon=0.016$.}
\label{fig:reversible_7}
\end{figure*}
\begin{figure*}
\centering
\includegraphics[scale=.46]{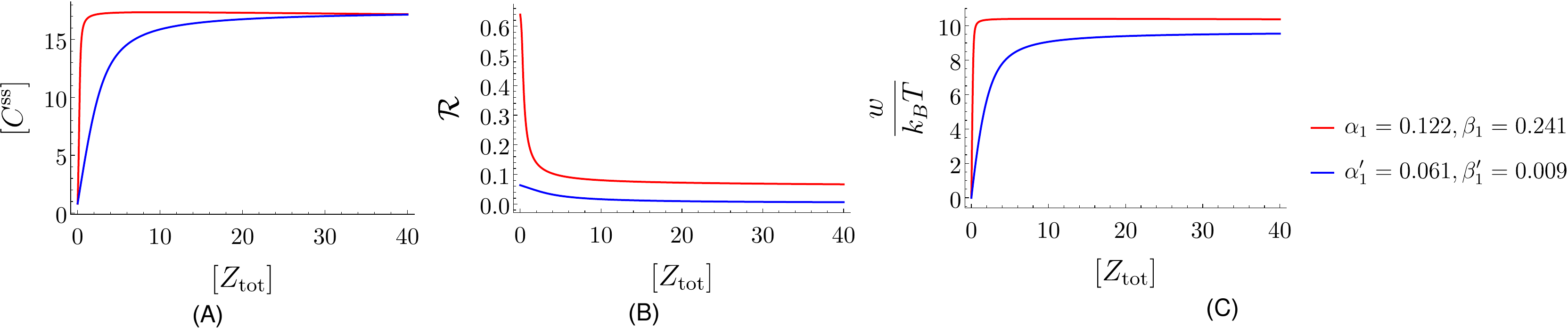}
\caption{$[X_{\text{tot}}]=58.564,[Y_{\text{tot}}]=4.7,[P_{\text{tot}}]=23.929,\alpha_1=0.122,\alpha_2=0.128,\beta_1=0.240,\beta_2=1.023,k_1=6.555,k_2=0.252,k_{\text{on}}=0.680,k_{\text{off}}=9.066,\epsilon=0.061$.}
\label{fig:reversible_8}
\end{figure*}
\begin{figure*}
\centering
\includegraphics[scale=.46]{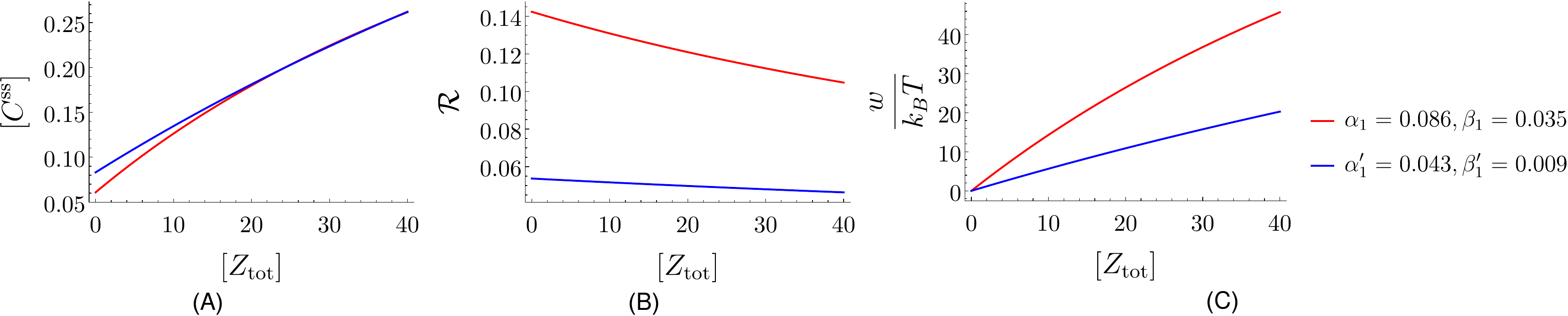}
\caption{$[X_{\text{tot}}]=18.954,[Y_{\text{tot}}]=166.734,[P_{\text{tot}}]=4.529,\alpha_1=0.086,\alpha_2=0.131,\beta_1=0.035,\beta_2=0.524,k_1=1.245,k_2=7.211,k_{\text{on}}=0.438,k_{\text{off}}=3.268,\epsilon=0.043$.}
\label{fig:reversible_9}
\end{figure*}
\begin{figure*}
\centering
\includegraphics[scale=.46]{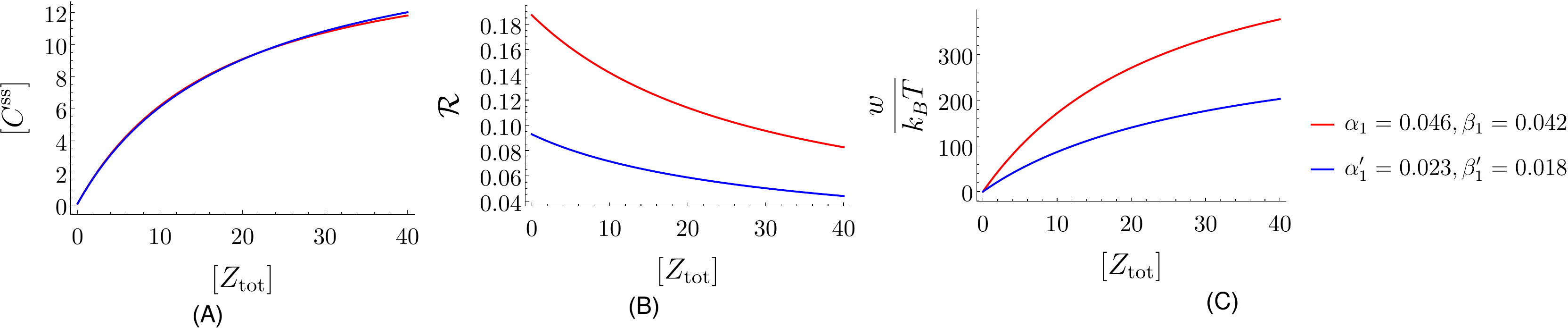}
\caption{$[X_{\text{tot}}]=54.201,[Y_{\text{tot}}]=50.055,[P_{\text{tot}}]=39.431,\alpha_1=0.045,\alpha_2=0.295,\beta_1=0.042,\beta_2=0.135,k_1=9.537,k_2=6.173,k_{\text{on}}=0.038,k_{\text{off}}=1.337,\epsilon=0.023$.}
\label{fig:reversible_10}
\end{figure*}
\end{multicols}

\end{document}